\documentclass[a4paper,preprint]{revtex4-1}
\usepackage[utf8]{inputenc}
\usepackage{graphicx}
\usepackage{mathtools}
\usepackage{amsfonts}
\usepackage{chemmacros}
\usepackage{multirow}
\usepackage{csquotes}
\usepackage{hyperref}

\usepackage[
	poorman,
	capitalize,
]{cleveref}

\begin{document}

\title{Analysis of Electron Correlation Effects in Strongly Correlated Systems (\ch{N2} and \ch{N2+}) by applying DMRG and Quantum Information Theory}

\author{Christian Stemmle}
\email{christian.stemmle@fu-berlin.de}
\author{Beate Paulus}
\affiliation{Institut f\"ur Chemie und Biochemie - Takustr. 3, 14195 Berlin, Freie Universit\"at Berlin, Germany}
\author{\"Ors Legeza}
\affiliation{Strongly correlated systems ``Lend\"ulet'' research group, Wigner Research Centre for Physics, P.O.Box 49 Hungary}

\begin{abstract}

	The dissociation of \ch{N2} and \ch{N2+} has been studied by using the \emph{ab initio} Density Matrix Renormalization Group (DMRG) method. Accurate Potential Energy Surfaces (PES) have been obtained for the electronic ground states of \ch{N2} ($\mathrm{X}\prescript{1}{}\Sigma_g^+$) and \mbox{\ch{N2+} ($\mathrm{X}\prescript{2}{}\Sigma_g^+$)} as well as for the \ch{N2+} excited state $\mathrm{B}\prescript{2}{}\Sigma_u^+$.
	Inherently to the DMRG approach, the eigenvalues of the reduced density matrix ($\rho$) and their correlation functions are at hand. Thus we can apply Quantum Information Theory (QIT) directly, and investigate how the wave function changes along the PES and depict differences between the different states.
	Moreover by characterizing quantum entanglement between different pairs of orbitals and analyzing the reduced density matrix, we achieved a better understanding of the multi-reference character featured by these systems.

\end{abstract}

\maketitle

\section{Introduction}

Accurate descriptions of the electronic structure for chemical systems are important for predicting molecular properties and reactivities. However, this requires numerically feasible methods for solving the electronic Schr\"odinger equation, which are difficult to obtain. A basic method for this task is the Hartree-Fock (HF) method, which represents the total electronic wave function as a single \emph{determinant} (configuration) build up as a product of \emph{one-electron wave functions} (orbitals). This mean-field approximation to the electron-electron interaction is computationally feasible, but introduces a systematic error, known as electron \emph{correlation}. A general and exact method, capable of correcting this error, is long known with the Full Configuration Interaction (FCI) approach. Unfortunately its factorial scaling with the size of the problem (i.e. the number of electrons) makes these calculations only feasible for very small systems. Systematic approximations, which are required for larger systems, result in a variety of methods with different advantages and disadvantages, making them only applicable for suitable systems.\\

In these systematic approximations one restricts the calculation to only include certain electron configurations (determinants or configuration state functions). \emph{Single-reference} methods (e.g. CISD, CCSD(T), Many-Body Perturbation Theory) \cite{Bartlett-1981} manage to describe the major part of the electron correlation by improving the wave function based on one reference configuration (usually the HF configuration). Systems inadequately described by these methods are called \emph{strongly correlated} and require \emph{multi-configurational} and subsequent \emph{multi-reference} methods (e.g. MCSCF, MR-CI, MR-CC) including much more determinants or configuration state functions (CSF) \cite{Pauncz-1979}.\\

To develop new efficient approaches for systematic approximations one can investigate and analyze the contribution of different configurations to the total electronic wave function.
Therefore these configurations may be assigned according to an (artificial) classification.
\citet{Bartlett-1994,Bartlett-2007} provided a classification into \emph{dynamic} (or \emph{weak}), \emph{static} (or \emph{strong}) and \emph{nondynamic} correlation. The dynamic correlation is subject to a large number of configurations, each with only a small contribution to the total wave function, and can be recovered by single-reference methods. The other two require multi-configuration approaches, and usually depend on a smaller number of configuration, which have a larger amplitude instead.
However, within this classification, there are no strict definitions for the different cases and some configurations may be assigned to multiple types. \\

The most popular approaches for dealing with multi-reference problems are the Complete Active Space Self-Consistent Field (CASSCF) method for static correlation \cite{Roos-1980,Szalay-2012} and the Multi-Reference Configuration Interaction (MR-CI) method for dynamic correlation on top of a multi-configurational wave function \cite{Szalay-2012}. The criterion to decide which configuration will be included in the wave function, depends on the choice for the active space and/or the restrictions to certain levels of excitations (Singles, Doubles, Triples, etc.). Thus the decision is biased and made before performing the actual calculation, possibly omitting unexpected, yet important, configurations.\\

A promising alternative approach to this is the Density Matrix Renormalization Group (DMRG) approach\cite{White-1992,White-1993,White-1999}, which tries to find the most important configurations during its iterative procedure, thus resulting in an unbiased truncation to the FCI wave function\cite{Legeza-2008,Chan-2008,Marti-2010,Wouters-2014,Kurashige-2014b,Szalay-2015}.
The advantage of DMRG is, that during this iterative procedure, the density matrix corresponding to the electronic wave function, is calculated. It can thus directly be used to analyze the different contributions to the wave function by applying Quantum Information Theory (QIT)\cite{Legeza-2003b,Legeza-2004b,Rissler-2006,Szalay-2015}. This gives us a measure of how important an orbital is for the different configurations required in the CI expansion. Studies using DMRG and QIT to investigate strongly correlated systems include Heisenberg spin chains \cite{Barcza-2015}, extended periodic Anderson model \cite{Hagymasi-2015}, graphene nanoribbons \cite{Hagymasi-2016a}, \ch{Be6} rings \cite{Fertitta-2014}, iron nitrosyl complexes \cite{Boguslawski-2012}, uranium carbide oxide \ch{CUO} \cite{Tecmer-2014}, photosystem II \cite{Kurashige-2013}, and Ru-NO bond in a Ruthenium nitrosyl complex \cite{Freitag-2015}. Additionally \citet{Boguslawski-2013} analyzed the dissociation of the electronic ground states of \ch{N2}, \ch{F2} and \ch{CsH} by comparing DMRG calculations with different active spaces.
Furthermore, QIT results for localized orbitals can be used as a tool to analyze the bonding character and identify covalent, dative or delocalized (e.g. aromatic) bonds\cite{Szilvasi-2015,Szalay-2017}.\\

A popular example for strongly correlated systems is the nitrogen molecule \ch{N2}. Especially when considering bond stretching up to dissociation, non-dynamic and static correlations become non-negligible.
The electronic structure of this molecule and its singly charged cation was already subject to numerous studies \cite{Roche-1975,Langhoff-1987,Langhoff-1988,Baltzer-1992,Chan-2004,Aoto-2006,Rissler-2006,Eckstein-2015,Paulus-2016}.
Here we want to study the ground states together with some selected excited states of \ch{N2} and \ch{N2+} by applying DMRG and QIT, and investigate the multi-reference character of the wave function with the ultimate aim of improving correlation methods.
The presented results include full Potential Energy Surfaces (PES) as well as eigenvalue spectra of reduced density matrices for a detailed analysis of electron correlation effects.
\\

\section{Density Matrix Renormalization Group (DMRG) and Quantum Information Theory (QIT)}
\label{sec:2_theory}
\begin{figure}
	\centering
	\includegraphics[width=0.8\linewidth]{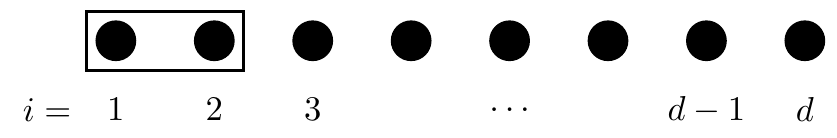}
	\caption{Schematic representation of the block chain. Each dot represents one orbital (site). The rectangle represents a sub-block, in this case it contains 2 orbitals, thus $4^2=16$ possible configurations can be formed within the sub-block.}
	\label{fig:block_chain}
\end{figure}
A detailed description of Density Matrix Renormalization Group (DMRG) and Quantum Information Theory (QIT) may be found various reviews by \citet{Legeza-2008,Marti-2010,Kurashige-2013,Wouters-2014,Szalay-2015,Olivares-Amaya-2015,Chan-2016}.
In the quantum chemistry version of the DMRG a one dimensional tensor topology is formed from molecular orbitals as shown in \cref{fig:block_chain}. More complex tensor networks like in the tree-tensor-network state (TTNS) algorithm are possible as well \cite{Murg-2010,Nakatani-2013,Murg-2015,Szalay-2015}. The CI wave function can then be written as
\begin{equation}
	|\Psi\rangle = \sum_{\alpha_1,\dots,\alpha_d} U(\alpha_1,\dots,\alpha_d) |\phi_{\alpha_1}^{\{1\}}\rangle\otimes\dots\otimes|\phi_{\alpha_d}^{\{d\}}\rangle
\end{equation}
where $\alpha_i$ labels the $d$ single-orbital basis states $|\phi_{\alpha_i}^{\{i\}}\rangle$ with the superscript ${\{i\}}$ indicating the orbitals position in the block chain and $U(\alpha_1,\dots,\alpha_d)$ are the coefficients arranged in a tensor of order $d$.
Each index $\alpha_i$ goes over the $q=4$ spin occupations of a spatial orbital (or ``single-orbital basis states''): $|\phi^{(1)}_\alpha\rangle \equiv |-\rangle$,  $|\phi^{(2)}_\alpha\rangle \equiv |\downarrow\rangle$,  $|\phi^{(3)}_\alpha\rangle \equiv |\uparrow\rangle$ and $|\phi^{(4)}_\alpha\rangle \equiv |\uparrow\downarrow\rangle$. Thus $U(\alpha_1,\dots,\alpha_d)$ has $q^d$ coefficients in total. 
Note, that these configurations can have any numbers of electrons from 0 to $2d$.\\

Since the memory requirements of the $d$-order tensor $U(\alpha_1,\dots,\alpha_d)$ grows exponentially with the number of orbitals $d$, it is required to factorize it as a product of low order tensors and with
controlled rank. The simplest case is the so-called Matrix Product State (MPS) representation, where
\begin{equation}
	U(\alpha_1,\dots,\alpha_d) = \boldsymbol{A}_1(\alpha_1)\boldsymbol{A}_2(\alpha_2)\cdots\boldsymbol{A}_{d-1}(\alpha_{d-1})\boldsymbol{A}_d(\alpha_d).
\end{equation}\\

Each matrix $\boldsymbol{A}_i(\alpha_i)$ thus corresponds to one molecular orbital (or site).
Note, that the size of the matrices still grows exponentially with increasing system size \cite{Schollwock-2005}, thus the MPS itself does not reduce the memory requirements. Instead we can define an upper limit to the matrix dimensions called \emph{number of block states} or \emph{virtual bond dimensions} $M$. It is, however, a non-trivial procedure how to choose a proper $M$ value. \\

In practice, the DMRG method provides an optimized set of $\boldsymbol{A}_i(\alpha_i)$ matrices.
The quantum correlations are taken into account by an iterative
procedure that  variationally minimizes the energy of
the Hamiltonian. 
The method converges to the full Configuration Interaction (CI) solution within the selected active orbital space.
In the two-site DMRG variant \cite{White-1992,Schollwock-2005},
the Hilbert space of $N_{\rm e}$ electrons correlated on $d$ orbitals, $\Lambda^{(d)}$,  is approximated by a tensor product
space of four tensor spaces defined on an ordered orbital chain, i.e.,
$\Xi^{(d)}_{\rm DMRG}=\Xi^{(\mathrm{l})}\otimes\Lambda_{i+1}\otimes\Lambda_{i+2}\otimes\Xi^{(\mathrm{r})}$.
The basis states of the $\Xi^{(\mathrm{l})}$
comprises $i$ orbitals to the left of the chain ($l \equiv$ left ) and  $\Xi^{(\mathrm{r})}$ comprises $d-i-2$ orbitals
to the right of the chain ($r \equiv$ right). These states
are determined  through a series of unitary transformations based on
the singular value decomposition (SVD) theorem
by going through the ordered orbital space from {\it left} to {\it right} and then sweeping back and forth \cite{Schollwock-2005,Szalay-2015}.
The number of block states, $M_l={\rm \dim}\ \Xi^{(\mathrm{l})}$
and $M_r={\rm \dim}\ \Xi^{(\mathrm{r})}$, required to achieve
sufficient convergence can be regarded as a function of the level of
entanglement among the orbitals \cite{Vidal-2003a,Legeza-2003b}.
The maximum number of block states
$M_{\rm max} = \max{(M_l,M_r)}$\ required to reach an
\emph{a priory} defined  accuracy threshold,  is inherently determined by
truncation error, $\delta \varepsilon_{\rm TR}$,
when the Dynamic Block State Selection (DBSS) approach is
used \cite{Legeza-2003a,Legeza-2004b}.
During the initial sweeps of the DMRG algorithm the
accuracy is also influenced by the
environmental error, $\delta \varepsilon_{\rm sweep}$\cite{Legeza-1996}.
The latter error can be reduced significantly by taking advantage of
the CI based  Dynamically  Extended Active Space
procedure (CI-DEAS) \cite{Legeza-2003b,Legeza-2004a}
and using a large number of DMRG sweeps
until the energy change between
two sweeps is negligible.
In the CI-DEAS procedure the active space of orbitals is extended
dynamically based on the orbital entropy
profile \cite{Barcza-2011,Szalay-2015}.
$M_{\rm max}$
depends strongly on the orbital ordering along the
one-dimensional chain topology of the DMRG
method \cite{Chan-2002a,Legeza-2003b,Moritz-2005,Barcza-2011}.
There  exist various extrapolation schemes to determine the
truncation-free solution \cite{Legeza-1996,Wouters-2014}.\\

To analyze the CI wave functions by means of QIT the $n$-orbital density matrix is needed. Formally it is obtained by a summation over all but $n$ orbitals, e.g. the \emph{one-orbital} density matrix is given by
\begin{widetext}
\begin{align}
	\rho_i(\alpha_i,\alpha_i^\prime) &= \mathrm{Tr}_{1,\dots,\not{i},\dots,d}|\Psi\rangle\langle\Psi| \\
	&= \sum_{\alpha_1,\dots,\not{\alpha_i},\dots,\alpha_d} U(\alpha_1,\dots,\alpha_i,\dots,\alpha_d) \overline {U(\alpha_1,\dots,\alpha_i^\prime,\dots,\alpha_d)}.
\end{align}
\end{widetext}
In DMRG this is quite easily obtained by contracting the MPS with itself over all indices except for $\alpha_i$ and $\alpha_i^\prime$.

One can quantify the contribution of an orbital $i$ to the correlation energy by means of the \emph{one-orbital} von Neumann entropy \cite{Legeza-2003b}
\begin{equation}
	S_i = -\mathrm{Tr}\left(\rho_{i}\ln\rho_{i}\right) = -\sum_\alpha \omega_{i,\alpha} \ln \omega_{i,\alpha},
\end{equation}
where $\omega_{i,\alpha}$ are the eigenvalues of the \emph{one-orbital} density matrix $\rho_i(\alpha,\alpha^\prime)$, and give the \emph{probability} amplitudes of the single-orbital spin states.
A small entropy is connected to small correlation effects of the corresponding orbital. Highest entropy is achieved when the 4 possible spin states are evenly distributed, i.e.~when $\omega_{i,\alpha}=0.25$ for all $\alpha$, then $S_i=-4\times0.25\times\ln0.25=\ln4\approx1.39$.
The sum of all one-orbital entropies gives a measure for the \emph{total correlation}
\begin{equation}
	I_{\mathrm{tot}} = \sum_i S_i
\end{equation}
of the wave function \cite{Legeza-2004b,Szalay-2017}.\\

Similarly the \emph{two-orbital} von Neumann entropy $S_{ij}$ is obtained from the \emph{two-orbital} density matrix $\rho_{ij}$ \cite{Legeza-2006a}.
\begin{align}
	\rho_i(\alpha_i,\alpha_j,\alpha_i^\prime,\alpha_j^\prime) &= \mathrm{Tr}_{1,\dots,\not{i},\dots,\not{j},\dots,d}|\Psi\rangle\langle\Psi| \\
	S_{ij} &= -\mathrm{Tr}\left(\rho_{ij}\ln\rho_{ij}\right) = -\sum_\alpha \omega_{ij,\alpha} \ln \omega_{ij,\alpha}.
\end{align}
It gives the contribution of two combined orbitals to the correlation energy.
If both orbitals are not correlated with each other, the two-orbital entropy becomes the sum of both single-orbital entropies. Any correlation between these two orbitals reduces the entropy of the two combined orbitals with the rest of the system, hence we can define the \emph{two-orbital mutual information} \cite{Rissler-2006} as
\begin{equation}
	I_{ij} = S_i + S_j - S_{ij}
	\label{eq:I_ij}
\end{equation}
describing the correlation of both classical and quantum origin between the two orbitals $i$ and $j$.\\

More detailed information is included in the \emph{eigenvalues} $\omega_{i,\alpha}$ of the one-orbital density matrix $\rho_{i}$, representing the \emph{probability} (amplitude) of the spin occupations $|\phi_\alpha\rangle$.
Similar, from the two-orbital density matrix $\rho_{ij}$ one obtains the \emph{eigenvalues} $\omega_{ij,\alpha}$ and \emph{eigenvectors} $\phi_{ij,\alpha}$, giving information about the spin probabilities of the orbital pair $ij$.
Note that $\alpha$ for the two-orbital density matrix $\rho_{ij}$ goes over $4\times4=16$ states, expressed in a basis obtained by combining the 4 possible spin states for each orbital ($|\phi_{\alpha_i}\rangle|\phi_{\alpha_j}\rangle = \{|-,-\rangle, |-,\downarrow\rangle, |\downarrow,-\rangle, \dots \}$).\\

These information are complemented by the \emph{generalized correlation functions}\cite{Rissler-2006,Barcza-2015,Fertitta-2014}, which tells us about the important excitations (or transitions) an orbital pair can do.
Consider the \emph{transition} between an initial state $|\phi_\alpha\rangle$ and final state $|\phi_{\alpha^\prime}\rangle$ of orbital $i$. The \emph{transition operators} are defined as
\begin{equation}
	T^{(m)} = |\phi_{\alpha^\prime}\rangle\langle\phi_\alpha| \hspace{20pt} \mathrm{for}\ m=1,\dots,q^2,
\end{equation}
where the index $m$ numbers the possible transitions between the $q=4$ states of each orbital (a convention for the numbering can be found in Ref. \cite{Szalay-2015}). As the transition operators act on a general $n$-orbital wave function, the operators are modified such that they only act on the $i$-th orbital
\begin{equation}
	T_i^{(m_i)} = \mathbb{I}\otimes\dots\otimes\mathbb{I}\otimes T^{(m_i)}\otimes\mathbb{I}\otimes\dots\otimes\mathbb{I}
\end{equation}
where $T^{(m_i)}$ is on the $i$-th position and $\mathbb{I}$ being the $q\times q$ identity matrix.
Combining two of these operators we can express the transition of electrons from orbital $i$ to $j$ and calculate the expectation value $\langle T_i^{(m_i)} T_j^{(m_j)} \rangle$, resulting in the \emph{generalized} correlation functions.
Note, that only some combinations of these operators ($m_i$ and $m_j$) result in non-zero values, due to electron number and spin conservation. For example, a transition from $|\downarrow\rangle$ to $|-\rangle$ in orbital $i$ would require the creation of a down-spin electron in orbital $j$ to conserve quantum numbers of the total electronic wave function (see Ref. \cite{Szalay-2015} for more details). \\

As a simple example consider two sets of configurations
\begin{align}
	|A\rangle &= |\phi_{\alpha_i},\phi_{\alpha_j},\phi_\beta\rangle = |\uparrow,\downarrow,\phi_\beta\rangle\\
	|B\rangle &= |\phi_{\alpha_i^\prime},\phi_{\alpha_j^\prime},\phi_\beta\rangle= |\downarrow,\uparrow,\phi_\beta\rangle = T_i^{(7)}T_j^{(10)}|A\rangle
\end{align}
which are part of the CI basis set. Here $|\phi_\beta\rangle$ represents the environment consisting of all orbitals excluding $i$ and $j$. The operator with $m=7$ corresponds to the spin flip from $|\uparrow\rangle$ to $|\downarrow\rangle$ (in orbital $i$) and the $m=10$ operator flips in the other direction (in orbital $j$). Their CI coefficients are $a(\beta)=U(\alpha_i,\alpha_j,\beta)$ and $b(\beta)=U(\alpha_i^\prime,\alpha_j^\prime,\beta)$ respectively. The generalized correlation function can then be expressed as
\begin{align}
	\langle T_i^{(7)} T_j^{(10)} \rangle &= \sum_{\alpha_i^\prime,\alpha_j^\prime,\beta}\sum_{\alpha_i,\alpha_j,\beta} b(\beta)^* a(\beta)
		\langle \downarrow,\uparrow,\phi_\beta | T_i^{(7)} T_j^{(10)} | \uparrow,\downarrow,\phi_\beta \rangle \label{eq:correl}\\
		&= \sum_\beta  b(\beta)^* a(\beta)
\end{align}
In \cref{eq:correl} all configurations not matching $\langle B|$ and $|A\rangle$, in the bra and ket vector respectively, will vanish when the double transition operator $T_i^{(7)} T_j^{(10)}$ acts on $|\Psi\rangle$.
Thus a generalized correlation function will vanish if $a(\beta)$ or $b(\beta)$ are orthogonal to each other. For diagonal transition operators, that leave the spin states unchanged, we obtain the summed amplitude of all matching configurations.
For a more general treatment see Ref. \cite{Barcza-2015}.\\

To determine the correlation between two subsystems, one has to consider the \emph{connected} part of the generalized correlation function
\begin{equation}
	\langle T_i^{(m_i)} T_j^{(m_j)} \rangle_C = \langle T_i^{(m_i)} T_j^{(m_j)} \rangle - \langle T_i^{(m_i)} \rangle\langle T_j^{(m_j)} \rangle
	\label{eq:correl_C}
\end{equation}
which is constructed similar to the mutual information in \cref{eq:I_ij}, i.e. the uncorrelated part is substracted.\\

In summary, the one-orbital quantities $S_i$ and $\omega_{i,\alpha}$ are obtained from the one-orbital density matrix and give information about the correlation and occupations of a single orbital. From the two-orbital density-matrix we obtain the quantities $S_{ij}$, $\omega_{ij,\alpha}$, $\phi_{ij,\alpha}$, $I_{ij}$ and $\langle T_i^{(m_i)} T_j^{(m_j)} \rangle_C$, telling us which orbitals are correlated with each other and which transitions (combination of two configurations) are most important.\\

Such concepts of quantum information theory, have already been applied successfully to spin and ultra-cold atomic systems \cite{Barcza-2015}, extended Anderson model \cite{Hagymasi-2015}, topological Kondo insulators \cite{Hagymasi-2016a}, graphene nanoribbons \cite{Hagymasi-2016b}, Be-rings \cite{Fertitta-2014} and diatomic chemical compounds \cite{Szilvasi-2015} in the ground state in order to reveal the entanglement structure and examine the spectrum of subsystem density matrices to understand the origin of entanglement.

\section{Model and Computational Details}
\begin{figure}
	\centering
	\includegraphics[width=\linewidth]{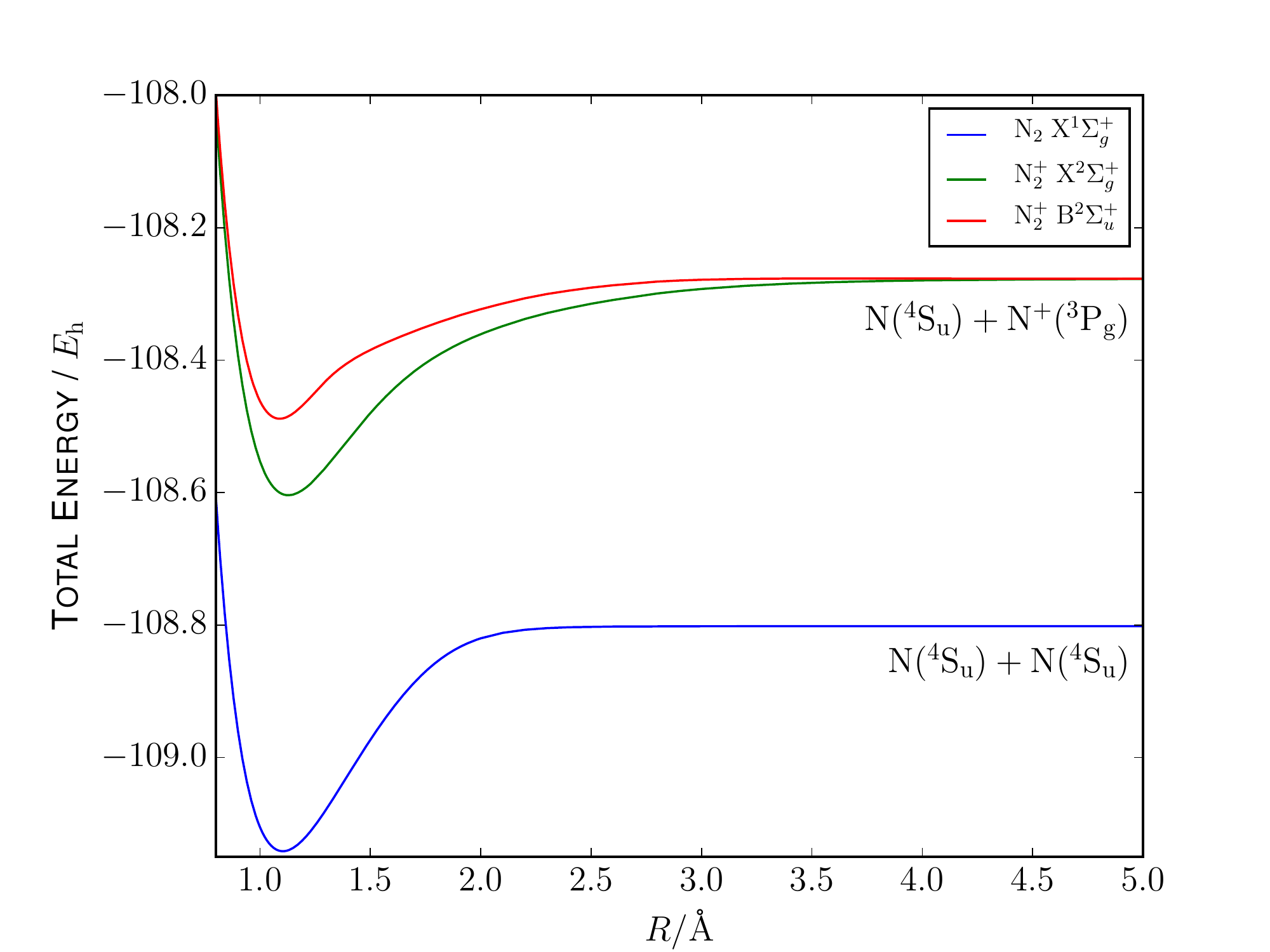}
	\caption{Potential Energy Surfaces (PES) for the electronic states considered in this work. 
	Results for CASSCF(9,8)/AV5Z and CASSCF(10,8)/AV5Z for \ch{N2+} and \ch{N2} respectively. The atomic states of the fragments are given for each dissociating limit.}
	\label{fig:pes_all_states}
\end{figure}
\begin{figure}
	\centering
	\includegraphics[height=0.8\textheight,width=\linewidth,keepaspectratio]{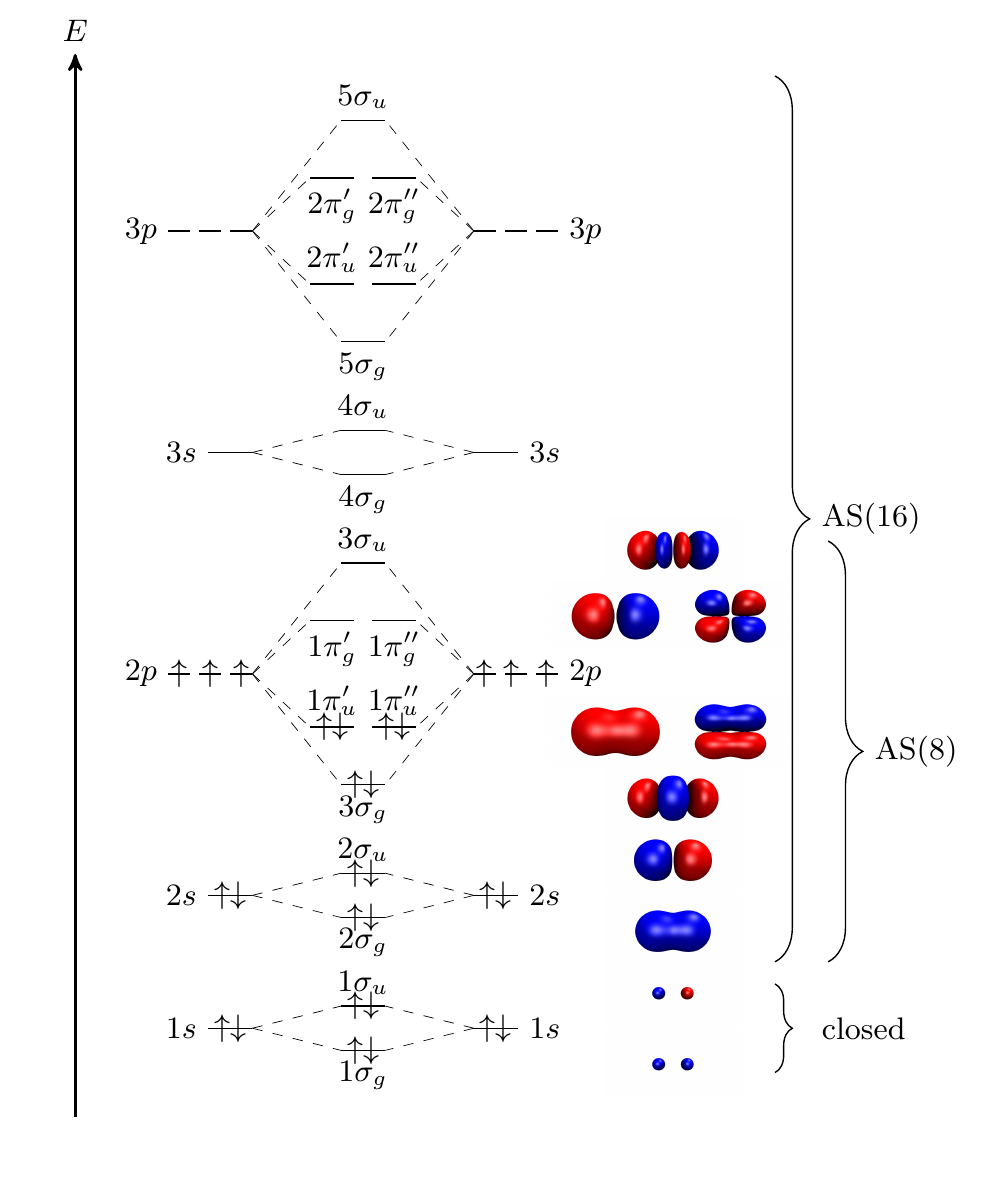}
	\caption{Schematic Molecular Orbital (MO) scheme for \ch{N2} with the occupation pattern of the electronic ground state configuration. Labels and plots of the orbitals as well as the used Active Spaces (AS), including 8 and 16 orbitals, are indicated.}
	\label{fig:moschema}
\end{figure}

To study the multi-reference character of the wave function we calculate and investigate the dissociation of \ch{N2} and \ch{N2+} for different electronic states of increasing complexity. For each state a PES will be calculated along the internuclear distance $R$ between the two nitrogen atoms.
We start with the closed-shell singlet electronic ground state of \ch{N2} labeled by the term symbol $\mathrm{X}\prescript{1}{}\Sigma_g^+$ and compare the results to the \ch{N2+} open-shell electronic ground state $\mathrm{X}\prescript{2}{}\Sigma_g^+$. Both states belong to the irreducible representation (IRREP) $A_g$ of the $D_{2h}$ point group.
Following this we extend the discussion to the higher excited \ch{N2+} state $\mathrm{B}\prescript{2}{}\Sigma_u^+$ with IRREP $B_{1u}$.
All regarded states are summarized in \cref{fig:pes_all_states}.\\

To obtain a suitable state average at the CASSCF level, we consider the different dissociation limits of the states of interest (see \cref{fig:pes_all_states}), and obtain all relevant electronic states by applying the Wigner-Witmer-Rules\cite{Herzberg-1950}. Further details for this specific example may also be found in a recent article by \citet{Liu-2014}.\\

Calculation are performed using \textsc{Molpro2012}\cite{Molpro2012} and applying the Dunning basis sets aug-cc-pVTZ (AVTZ), aug-cc-pVQZ (AVQZ) as well as aug-cc-pV5Z (AV5Z)\cite{Kendall-1992}. For the DMRG calculations the Budapest DMRG program \cite{DMRGcode} was used. The results are plotted using Python's Matplotlib library\cite{matplotlib}.\\

Two different active spaces are used, with 8 and 16 active orbitals respectively. In both cases the lowest lying core orbitals $1\sigma_g$ and $1\sigma_u$ are closed.
The choice of both active spaces is illustrated in \cref{fig:moschema}, together with the orbital labels. In the following we use a notation where the used method is appended by the size of the active space in parenthesizes (e.g. DMRG(16)). The number of electrons within these active orbitals is omitted, as it is always 10 in case of the neutral \ch{N2} molecule or 9 for the \ch{N2+} cation.\\

As a first step orbitals are optimized in a CASSCF(8) calculation including all states of same symmetry dissociating to the same asymptotic limit.
After transforming to natural orbitals, the required electron integrals for the DMRG calculation are exported to an integral file and include 16 active orbitals. Thus the DMRG calculations include 16 active orbitals (DMRG(16)), which corresponds to an CASCI(16) if the number of blockstates $M$ in the DMRG calculations are at the numerical exact limit.
As a reference for the DMRG calculations CASCI(16) calculations were performed, by requesting a MRCI calculation without external excitations (i.e. excitations into orbitals not part of the active space) were performed.
Furthermore we performed MRCI(8) calculations, to evaluate the effect of dynamical correlation.
In all cases MRCI refers to MRCI-SD without Davidson correction.
To illustrate the basis set effect, we present PES for the AVTZ, AVQZ as well as AV5Z basis sets in \cref{fig:pes_basis_methods}. All DMRG and QIT results are obtained using aug-cc-pV5Z (AV5Z).\\

The DMRG calculations have been performed in two runs. In the first run the orbitals are ordered as given by \textsc{Molpro} and a small number of blockstates $M=256$ is used, as this is sufficient for qualitative QIT results. Those can then be used to optimize the orbital ordering along the 1-dimension block chain according to the Fiedler vector\cite{Barcza-2011,Szalay-2015}. This ordering may change for different internuclear distances $R$. In a second run the number of blockstates is set to $M=4096$, being close to the numerical exact limit of CASCI(16).

\section{Results}
\label{sec:4_results}

We start the discussion by presenting the PES for the three states \ch{N2} $\mathrm{X}\prescript{1}{}\Sigma_g^+$, \mbox{\ch{N2+}} $\mathrm{X}\prescript{2}{}\Sigma_g^+$ and \ch{N2+} $\mathrm{B}\prescript{2}{}\Sigma_u^+$ in \cref{sec:4_PES}. We compare different methods and basis sets.\\

Next the orbitals are characterized in terms of atomic basis functions contribution and their energies in \cref{sec:4_orb_char}. These information will be helpful when analyzing the QIT results in the next sections, where we will see how strong each orbital is entangled by looking at the spin state probabilities $\omega_{i,\alpha}$ and orbital entropies $S_i$. Furthermore we present for selected pairs of orbitals the mutual information $I_{ij}$, the correlation functions $\langle T_i^{(m_i)} T_j^{(m_j)} \rangle_C$ as well as the eigenvalues and eigenvectors of the two-orbital density matrix $\rho_{ij}$ to investigate their correlation.
For simplicity we restrict the discussion here to the orbitals included in the AS(8), contributions form the AS(16) orbitals are much smaller in magnitude.\\

\subsection{Potential Energy Surfaces (PES)}
\label{sec:4_PES}
\begin{figure}
	\centering
	\includegraphics[width=\linewidth]{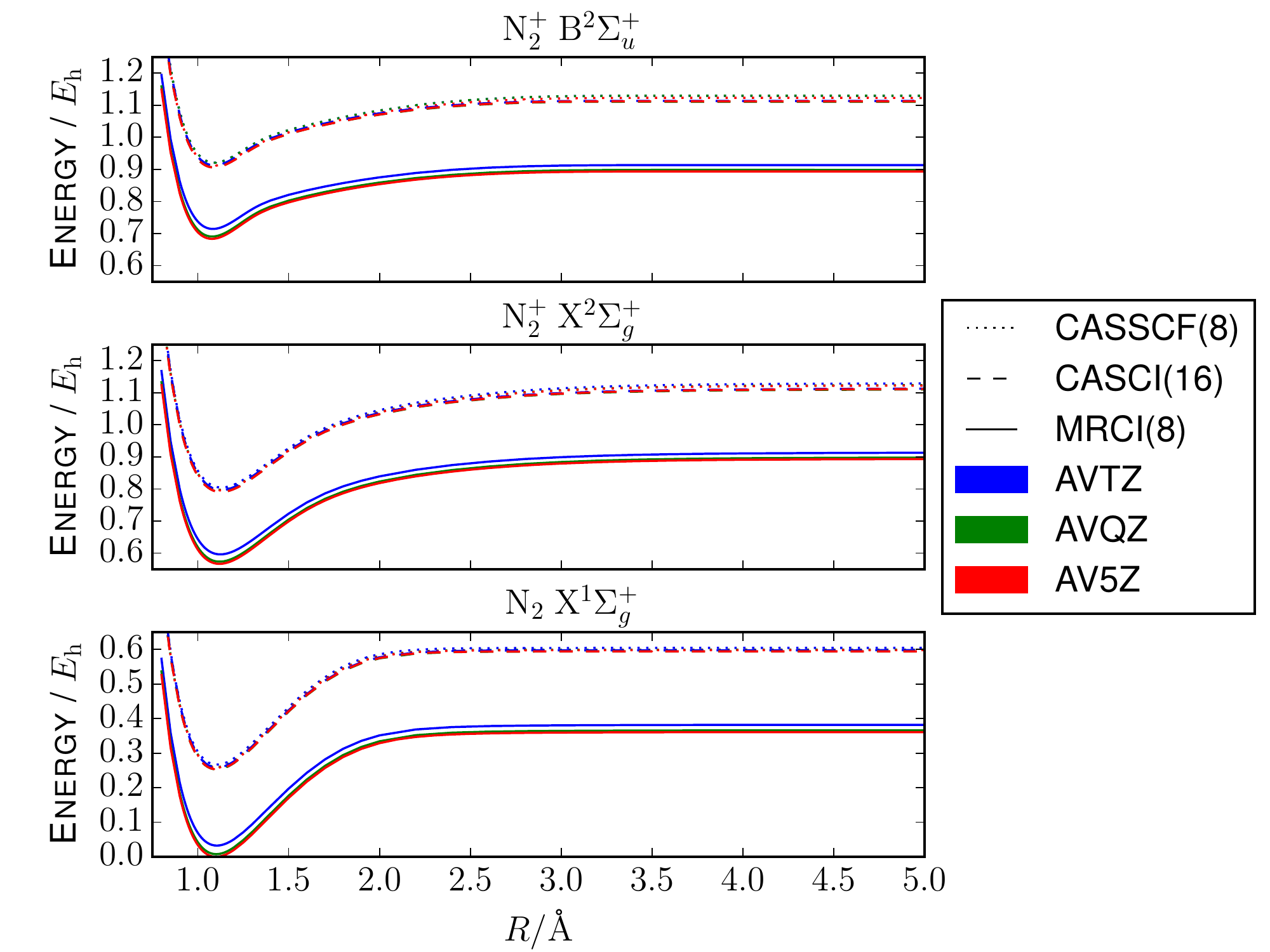}
	\caption{Reference Calculations for the Potential Energy Surfaces (PES) calculated with the different methods (CASSCF(8) (dotted), CASCI(16) (dashed) and MRCI(8) (continuous)) and basis sets (AVTZ (blue), AVQZ (green) and AV5Z(red)). Electronic states from top to bottom: \ch{N2+} $\mathrm{B}\prescript{2}{}\Sigma_u^+$, \ch{N2+} $\mathrm{X}\prescript{2}{}\Sigma_g^+$ and \ch{N2} $\mathrm{X}\prescript{1}{}\Sigma_g^+$.
	The AV5Z calculations are essentially at the complete basis set limit. Static correlation is slightly improved when choosing the larger active space ($\Delta E < 0.01\ E_{\rm h}$), while including dynamic correlation improves the energies by $0.2$ to $0.3\ E_{\rm h}$.
	The reference energy is set to the minimum of the \ch{N2} $\mathrm{X}\prescript{1}{}\Sigma_g^+$ at the CASCI(16)/A5VZ level.
	}
	\label{fig:pes_basis_methods}
\end{figure}
\begin{figure}
	\centering
	\includegraphics[width=\linewidth]{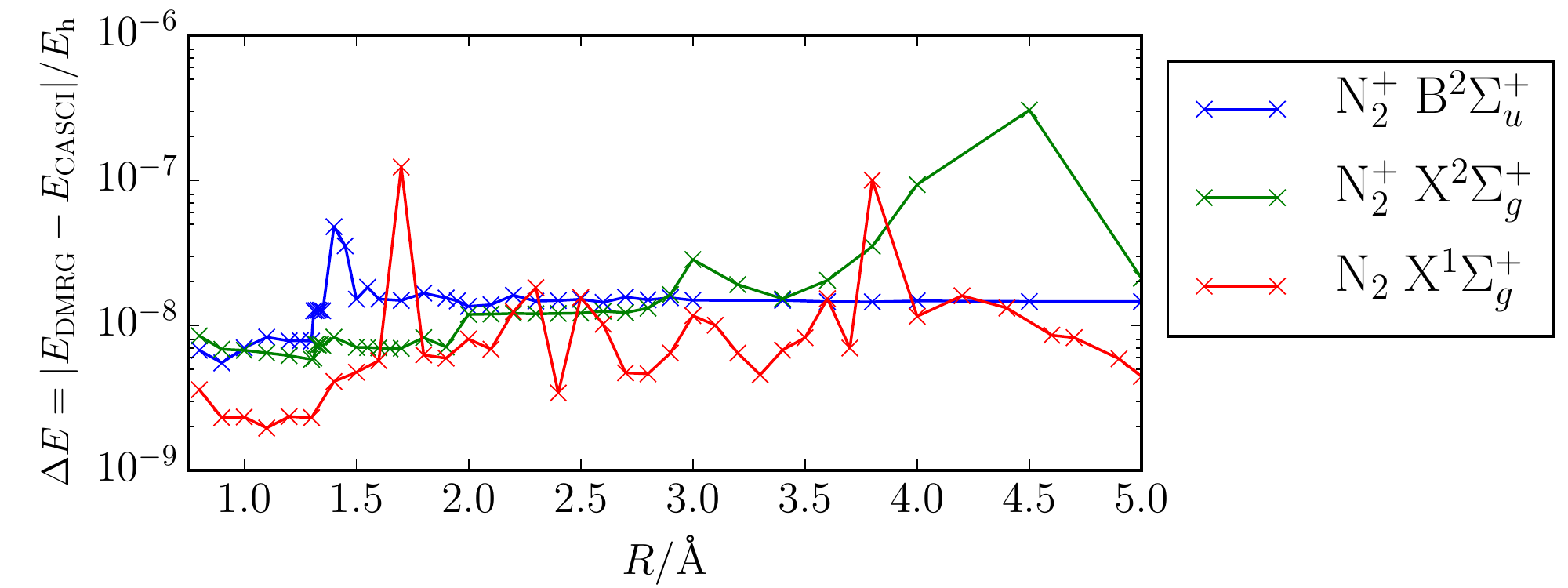}
	\caption{Energy difference between DMRG(16) and the CASCI(16) for all three states. With the largest value being smaller than $10^{-6} E_{\rm h}$, both methods are identical within numerical accuracy. DMRG calculations are performed using DBSS with a maximum limit for the blockstates of $M_{\rm max}=4096$ blockstates, which is close to the exact limit of CASCI(16).}
	\label{fig:pes_dmrg}
\end{figure}
\begin{figure}
	\centering
	\includegraphics[width=\linewidth]{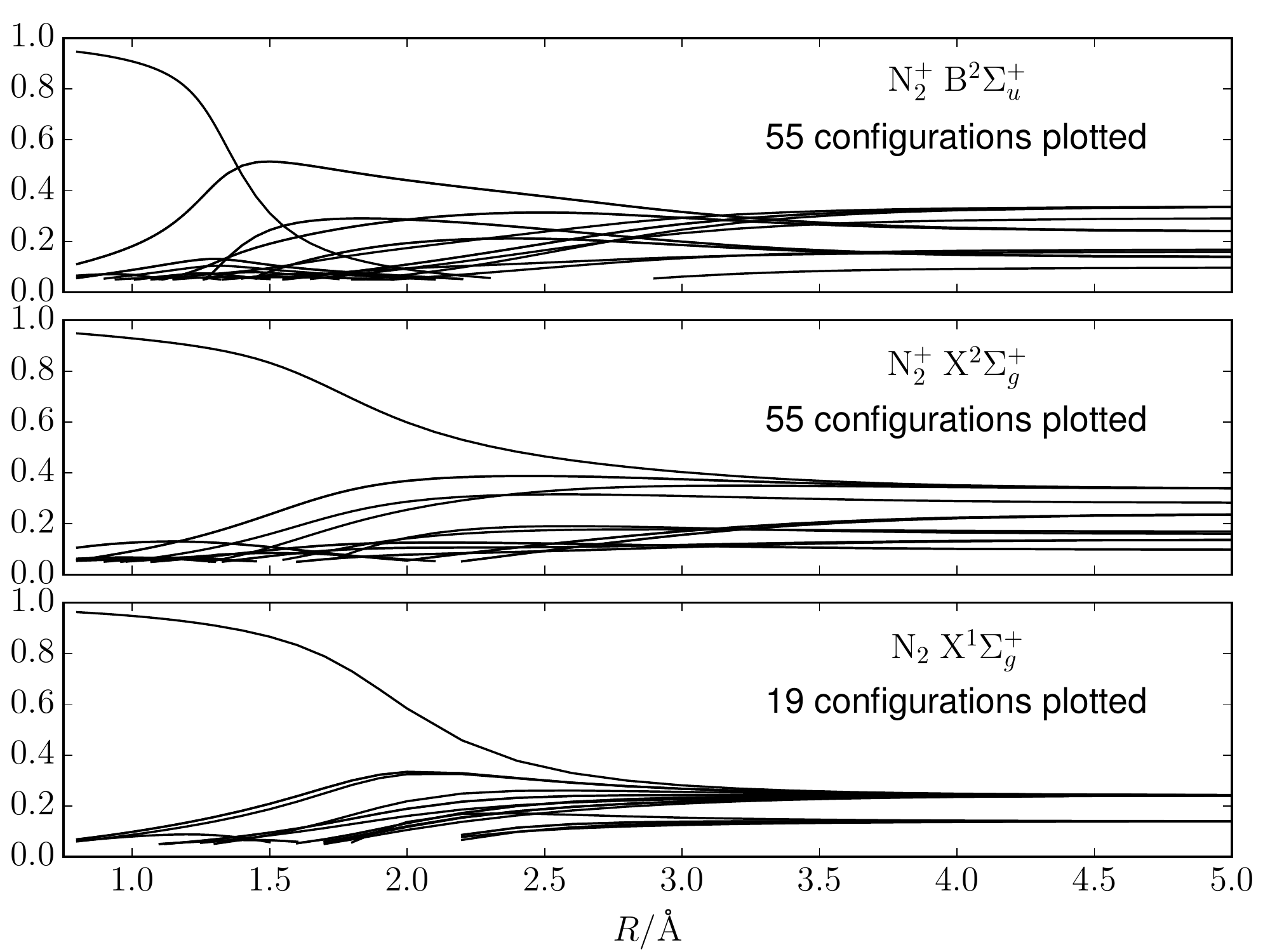}
	\caption{Coefficients of the most important configurations building up the CI wave function as obtained by CASSCF(9,8) and MRCI(8) calculations. In all cases the leading configuration has a very large coefficients around the equilibrium distance, but decreases when approaching the dissociation limit.}
	\label{fig:coeffs}
\end{figure}

\begin{table*}
	\centering
	\caption{Well Depths $D_e$ in Hartree ($E_{\rm h}$) for the different methods used. \newline \textsuperscript{a} calculated based on Dunham expansion and experimental values reported in ref \cite{NIST}}
	\label{tab:dissociation_energies}
	\begin{tabular}{llccccc}
		& & \textbf{CAS(8)} & \textbf{DMRG(16)} &  \textbf{CASCI(16)} & \textbf{MRCI(8)} & \textbf{lit.\textsuperscript{a}} \\
		\hline\hline
		\textbf{\ch{N2+}} $\mathbf{B\prescript{2}{}\Sigma_u^+}$ & AVTZ & 0.2088 &  & 0.2007 & 0.1990 & \multirow{3}{*}{} \\
			& AVQZ & 0.2088 &  & 0.2045 & 0.2076 &  \\
			& AV5Z & 0.2114 & 0.2059 & 0.2062 & 0.2101 & \\
		\hline
		\textbf{\ch{N2+}} $\mathbf{X\prescript{2}{}\Sigma_g^+}$ & AVTZ & 0.3244 &  & 0.3159 & 0.3167 & \multirow{3}{*}{$\Bigg\}$ 0.3256} \\
			& AVQZ & 0.3262 &  & 0.3193 & 0.3243 & \\
			& AV5Z & 0.3264 & 0.3200 & 0.3208 & 0.3265  & \\
		\hline
		\textbf{\ch{N2}} $\mathbf{X\prescript{1}{}\Sigma_g^+}$ & AVTZ & 0.3379 &  & 0.3396 & 0.3493 & \multirow{3}{*}{$\Bigg\}$ 0.3638} \\
			& AVQZ & 0.3397 &  & 0.3416 & 0.3581 & \\
			& AV5Z & 0.3399 & 0.3411 & 0.3411 & 0.3608 & \\
	\end{tabular}
\end{table*}

First, let us consider the PESs. In \cref{fig:pes_basis_methods} the electronic state are compared for different methods. In all three cases (from top to bottom \ch{N2+} $\mathrm{B}\prescript{2}{}\Sigma_u^+$, \ch{N2+} $\mathrm{X}\prescript{2}{}\Sigma_g^+$ and \ch{N2} $\mathrm{X}\prescript{1}{}\Sigma_g^+$) the picture is very similar: The energy difference between AVQZ and AV5Z is negligible, thus AV5Z is close to the complete basis set limit. The description of static correlation is only slightly improved when including more orbitals in the active space. The energy differences between CASSCF(8) and CASCI(16) are below $0.01\ E_{\rm h}$. Major improvements in the energy are obtained by including dynamical correlation on the MRCI(8) level, yielding energy differences up to $0.02\ E_{\rm h}$.\\

In \cref{fig:pes_dmrg} the DMRG(16) PES is compared to the CASCI(16) energies by showing their energy difference $\Delta E = E_{\rm DMRG} - E_{\rm CASCI}$.
The two methods yield the same energies (within numerical accuracy). They are thus in perfect agreement, as expected for a large enough number of blockstates $M$. Here, the DMRG(16) would be identical (i.e.~include exactly the same configurations) to CASCI(16) for a symmetric superblock configuration with $M=4^7=16384$ by partitioning seven orbitals to the left block and seven orbitals to the right block according to Fig.~\ref{fig:block_chain}.
Note, that the complete tensor $U(\alpha_1,\dots,\alpha_d)$ stores $4^{16}\approx4.3\times10^9$ configurations out of which only $\binom{16}{10}=8008$ and $\binom{16}{9}=11440$ are of interest for \ch{N2} and \ch{N2+}, respectively, due to conservation of total quantum numbers.
Using the DBSS approach with a density matrix truncation limit of $10^{-8}$, the number of blockstates is limited to values below 4096.\\

Though the main features of the PESs can be obtained by all methods, the well depths obtained with different methods show deviations (up to $0.02\ E_{\rm h}$) as summarized in \cref{tab:dissociation_energies}. Best agreement with literature values is obtained on the MRCI(8)/AV5Z level.\\

The coefficients of the CI wave function for the CAS(8) and MRCI(8) methods are plotted in \cref{fig:coeffs}, where we restrict ourselves to the most important ones, i.e. those where $|c_i|>0.05$ for any point along $R$. All pictures are rather similar, the leading configuration dominates the CI wave function around the equilibrium distance, then decreases with varying slopes. At the dissociation limit no single leading configuration can be determined, expressing the multi-reference character of the system.

\subsection{Orbital Characterization}
\label{sec:4_orb_char}
\begin{figure}
	\centering
	\includegraphics[width=0.7\linewidth]{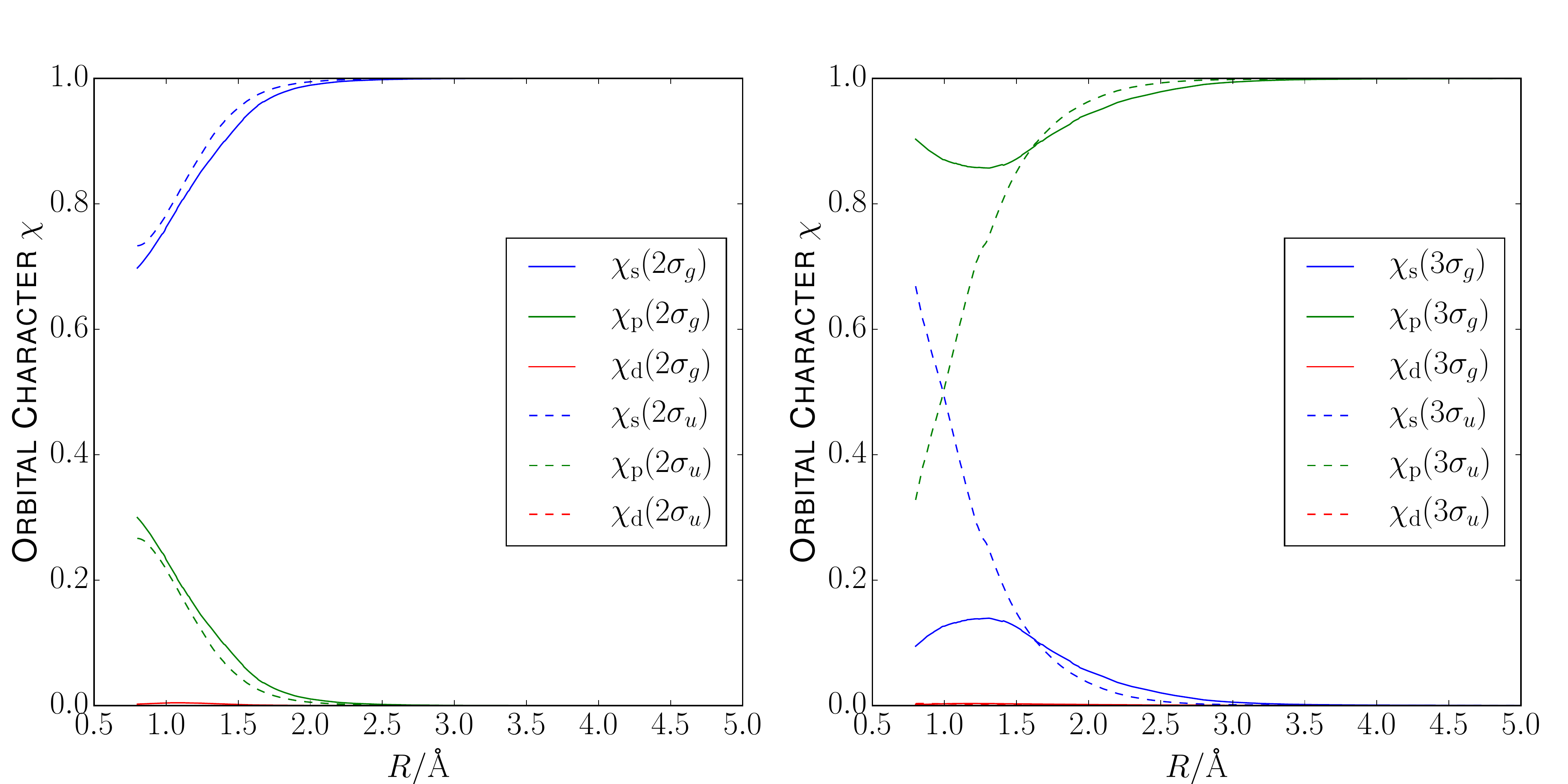}
	\includegraphics[width=0.7\linewidth]{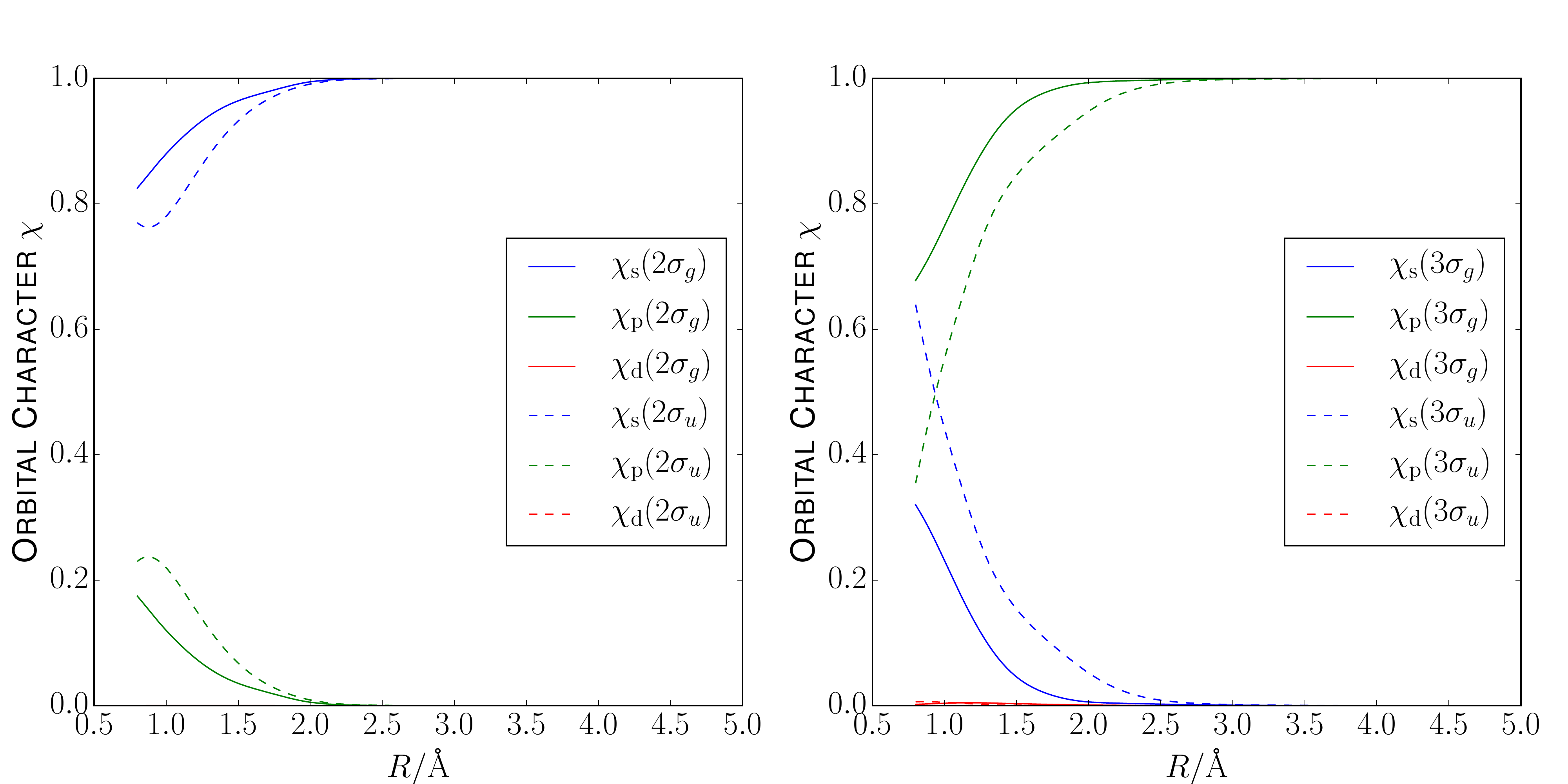}
	\caption{Character of the molecular orbitals for \ch{N2} (lower panels) and \ch{N2+} (upper panels) in terms of contribution of the atomic $s$, $p$ and $d$ orbitals as a function of the internuclear distance $R$. Only the most important $\sigma$ orbitals are shown, the $\pi$ orbitals have $\chi_p(\pi) \approx 1$. Orbitals optimized on the CAS(10,8)/AVQZ level.
	}
	\label{fig:orbchar}
\end{figure}
\begin{figure}
	\centering
	\includegraphics[width=0.49\linewidth]{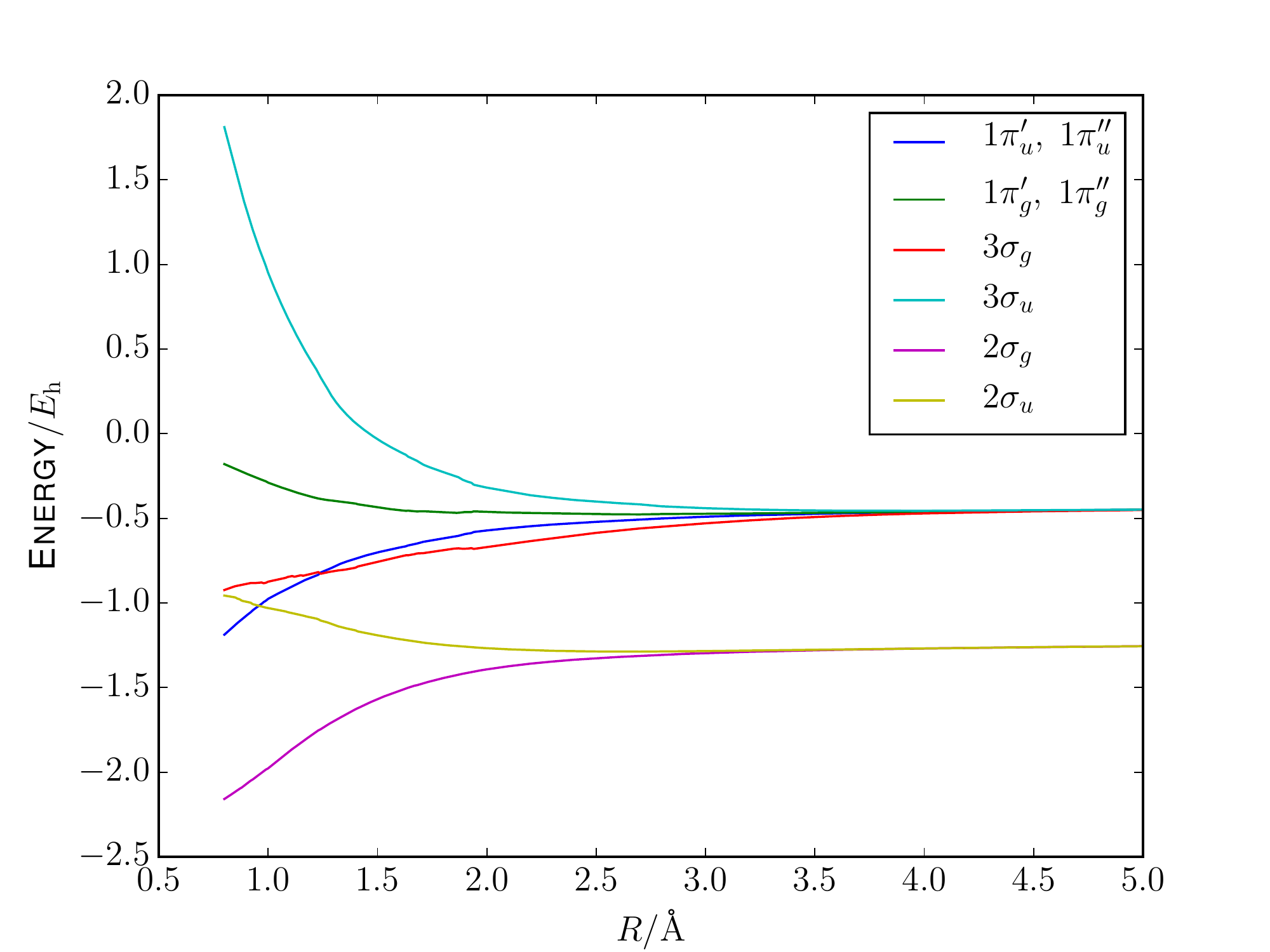}
	\includegraphics[width=0.49\linewidth]{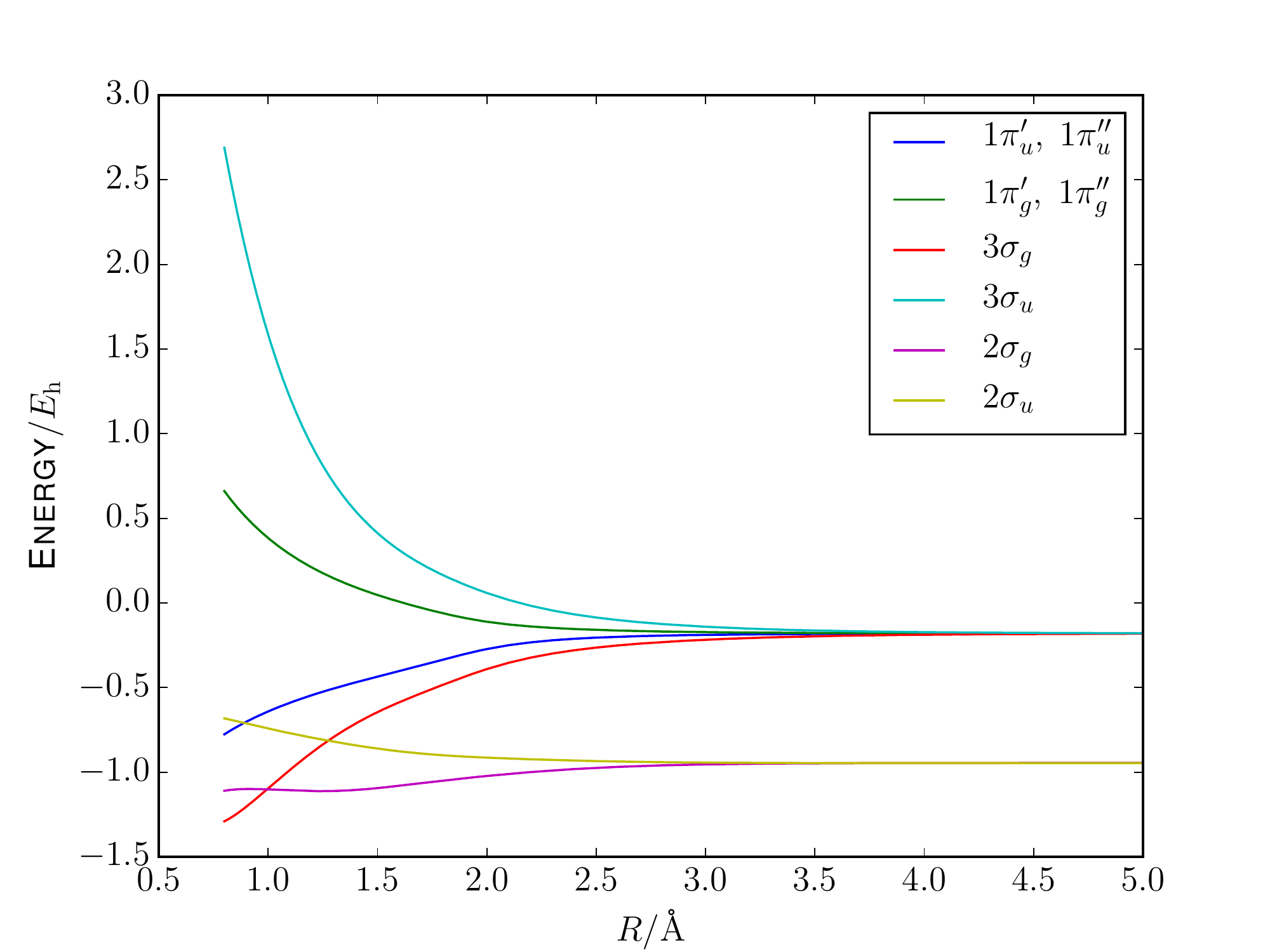}
	\caption{Energies of the molecular orbitals for the \ch{N2} (right) and \ch{N2+} (left). Orbitals optimized on the CAS(10,8)/AVQZ level.
	}
	\label{fig:orbenergy}
\end{figure}

Before investigating the entanglement patterns of the orbitals, it is helpful to characterize the orbitals in terms of their contributing atomic basis functions (i.e. $s$, $p$ and $d$ character) as well as energies. As the orbital energies and basis function character are very similar for the different electronic states we restrict the discussion here to the \ch{N2} ground state $\mathrm{X}\prescript{1}{}\Sigma_g^+$. The occupation patterns will be discussed for the other two states (\ch{N2+} $\mathrm{X}\prescript{2}{}\Sigma_g^+$ and \ch{N2+} $\mathrm{B}\prescript{2}{}\Sigma_u^+$) as well.\\

The atomic basis function character of the molecular orbitals is obtained by summation over the contributions of each type
\begin{equation}
	\chi_s = \sum_i |c_{s,i}|^2 \ ,\hspace{15pt} \chi_p = \sum_i |c_{p,i}|^2 \ ,\hspace{15pt} \chi_d = \sum_i |c_{d,i}|^2
\end{equation}
and renormalization to the constraint
\begin{equation}
	\chi_{\rm tot} = \chi_s + \chi_p + \chi_d = 1.
\end{equation}
It is evident from \cref{fig:orbchar}, that the $2\sigma_g$ and $2\sigma_u$ are dominated by $s$ character which increases with the internuclear distance $R$ until the molecular orbitals have evolved into the atomic $2s$ orbitals of each fragment at the dissociation limit. Similar the $3\sigma_g$ and $3\sigma_u$ are dominated by $p$ character and evolve into $p_z$ orbitals at the dissociation limit (cf. \cref{fig:moschema}).
The $3\sigma_u$ starts with a larger $\chi_s$ contribution than the $3\sigma_g$, but the behavior at the dissociation limit remains the same for both, as they become degenerate.
The different $\pi$ orbitals are not shown, as their $p$ contribution $\chi_p$ is virtually 1 for all internuclear distances due to symmetry constraints.\\

The orbital energies presented in \cref{fig:orbenergy} nicely show how all pairs of gerade and ungerade molecular orbitals largely differ at small internuclear distances and are slowly evolving to degenerate orbitals at the dissociate limit. Additionally to the gerade-ungerade-pairs, the $3\sigma$ orbitals become degenerate with the $1\pi$ orbitals at the dissociation limit, as they represent the three fold degenerate atomic $2p$ orbitals.\\

\subsection{Entanglement and Correlation of Single Orbitals and Orbital Pairs}

\begin{figure}
	\centering
	\includegraphics[width=\linewidth]{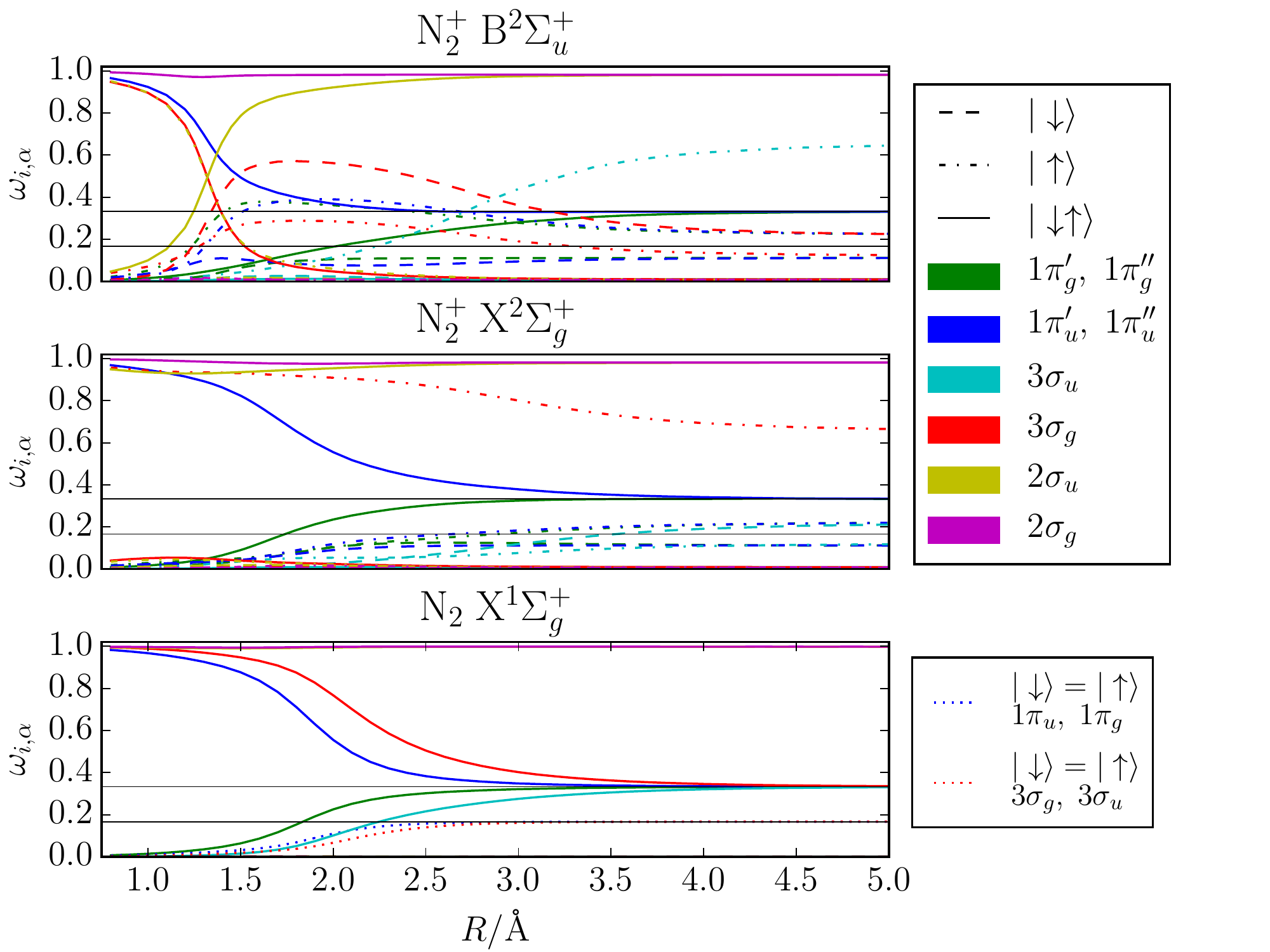}
	\caption{
		Eigenvalues of the one-orbital density matrix as a function of the internuclear distance $R$, representing the amplitude of the different spin occupation probabilities (distinguished by line styles) of each spatial orbital (distinguished by different colors). For simplicity, the values for the empty spin occupations $|-\rangle$ are omitted.
		For the singlet state in the bottom panel ($\mathrm{X}\prescript{1}{}\Sigma_g^+$), up-spin occupations ($|\uparrow\rangle$) are omitted as they are degenerate to their corresponding down-spin occupations ($|\downarrow\rangle$).
		The dotted lines represent coinciding curves, as indicated by the box in the lower right.
		The sum over all spin states for each orbital is always \mbox{$|-\rangle + |\downarrow\rangle + |\uparrow\rangle + |\uparrow\downarrow\rangle = 1$}.
		The two horizontal lines in each plot mark \nicefrac{1}{3} and \nicefrac{1}{6}.
	}
	\label{fig:w1}
\end{figure}
\begin{figure}
	\centering
	\includegraphics[width=\linewidth]{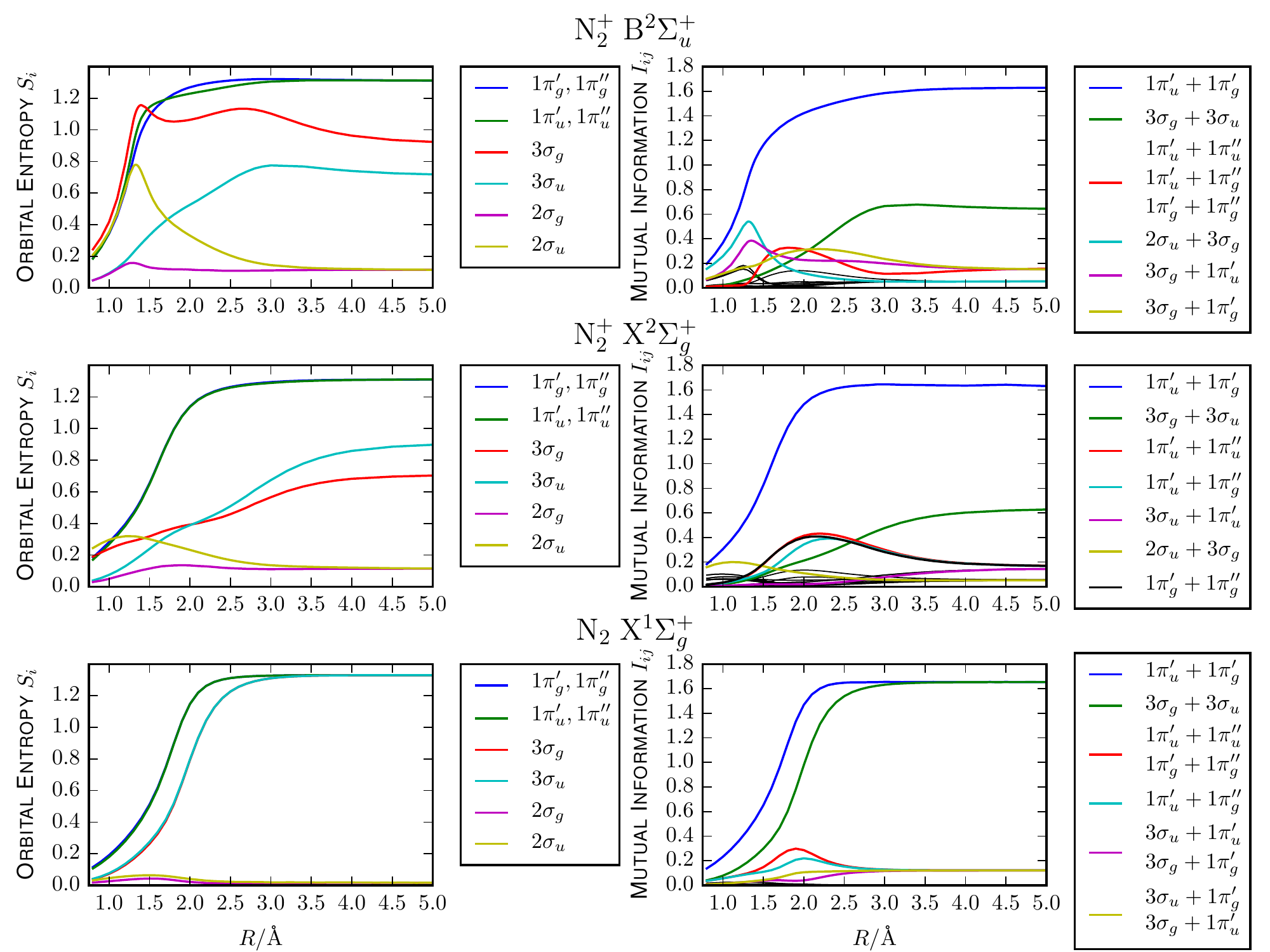}
	\caption{One orbital entropy $S_i$ (left column) and mutual information $I_{ij}$ (right column) for the most important orbitals of the three lowest states (from bottom to top: \ch{N2} $\mathrm{X}\prescript{1}{}\Sigma_g^+$, \ch{N2+} $\mathrm{X}\prescript{2}{}\Sigma_g^+$ and \ch{N2+} $\mathrm{B}\prescript{2}{}\Sigma_u^+$). Mutual Informations $I_{ij}$ of the considered orbitals pairs are printed in color, black lines indicate orbital pairs with less entanglement and are not further specified. Information that is redundant due to degeneracies is omitted.}
	\label{fig:S1_I}
\end{figure}

Let us now investigate the entanglement and correlation effects of orbitals, by analyzing the QIT quantities introduced in \cref{sec:2_theory}. We will discuss the three selected electronic states one by one.
Combined for all three states, the spin occupation probabilities $\omega_{i,\alpha}$ are shown in \cref{fig:w1}, while the single-orbital entropy $S_i$ and mutual information $I_{ij}$ between pairs of orbitals are shown in \cref{fig:S1_I}. For two selected orbital pairs, the diagonalized 2-orbital reduced density matrices and the connected part of the generalized correlation functions are presented in \cref{fig:phi2} and \cref{fig:correl} respectively. For the latter two quantities further orbital pairs are presented in the Supplementary Information (Figs. S1 to S5). \\

\subsubsection[N2 ground state]{\ch{N2} $\mathrm{X}\prescript{1}{}\Sigma_g^+$ ground state}

Let us start with the simplest case, the \ch{N2} $\mathrm{X}\prescript{1}{}\Sigma_g^+$ ground state. We will first look at the spin state probability in the lower plot of \cref{fig:w1}. Since the \ch{N2} ground state is a spin singlet state, the spin occupations for up-spin and down-spin are identical for all orbitals (dotted lines).
Due to degeneracy the spin occupations for $1\pi_{g/u}^\prime$ and $1\pi_{g/u}^{\prime\prime}$ coincide as well.
Additionally the $1\pi$ and $3\sigma$ orbitals converge towards the same two common dissociation limits, where we have in total a 6-fold degeneracy corresponding to the $2\times 3=6$ $2p$ atomic orbitals of the two $\mathrm{N} (\mathrm{\prescript{4}{}S}_u)$ fragments.
The two asymptotic limits are \nicefrac{1}{3} and \nicefrac{1}{6} for doubly and singly occupied states respectively and indicated by black horizontal lines. Adding together, we obtain for each spin (up or down) $\nicefrac{1}{6}+\nicefrac{1}{3}=\nicefrac{1}{2}$ probability for the $2p$ orbitals of the separated fragments.
On the other hand, around equilibrium distance, we can observe high double occupations for the bonding orbitals, and close to zero occupations for the anti-bonding orbitals. This simply reflects the occupation pattern of the leading HF configuration  (cf. \cref{fig:coeffs}).
Furthermore, while the $3\sigma$ and $1\pi$ orbitals change drastically with increasing internuclear distance $R$, the $2\sigma_g$ and $2\sigma_u$ occupations remain close to the full and empty spin states respectively. Hence their contribution to the correlation energy will be small and rather constant with respect to $R$.
Finally, comparing the $3\sigma$ and $1\pi$ orbitals, we observe the convergence of the $3\sigma$ orbitals to be shifted to slightly larger internuclear distances. This effect can be explained with the larger overlap of the $p_z$ orbitals as they are aligned along the molecular axis. Thus the $\pi$ bonds are broken sooner than the $\sigma$ bond. This was already previously recognized and described by \citet{Boguslawski-2013} for the one-orbital entropy.\\

After analyzing the occupation numbers, the one-orbital entropy (lower left plot in \cref{fig:S1_I}) can be easily understood. We observe only small entropies at the equilibrium distance, where the HF configuration captures most of the electron-electron interaction.
At the dissociation limit, where we have more electron correlation, the entropies achieve values close to their maximum of $\ln 4 \approx 1.39$ for the $3\sigma$ and $1\pi$ orbitals, while the $2\sigma$ entropies start small and decrease even further.
Additionally, just as for the orbital occupations $\omega_{i,\alpha}$, we observe the $3\sigma$ curves to be shifted towards larger bond lengths (when compared to the $1\pi$ ones). This again indicates the $\pi$ bonds to be broken before the $\sigma$ bond \cite{Boguslawski-2013}.\\

Furthermore, the $3\sigma_g$ and $3\sigma_u$ entropies are very similar, with small deviations for small $R$. A closer look at the occupations reveals their occupations to be nearly symmetric, i.e. the up and down-spin occupations are almost identical (in fact there are very small deviations for small $R$, but those are neglected in \cref{fig:w1} for a better visibility). Similarly, the double occupation of $3\sigma_g$ is almost identical to the empty occupation of $3\sigma_u$ (again not shown explicitly). Consequently the orbital entropies are (almost) identical. A similar discussion can be conducted for the $1\pi_{g/u}$ orbitals.\\

The single orbital picture only tells us, which orbitals are highly correlated. If we also wish to understand with what other orbitals they are mainly entangled, we can analyze the orbitals pairs in terms of the mutual information $I_{ij}$, eigenvalues $\omega_{ij}$ and corresponding eigenvectors $\phi_{ij}$ of the diagonalized 2 orbital reduced density matrices $\rho_{ij}$ as well as the connected contributions of the generalized correlation functions $\langle T_i^{(m_i)} T_j^{(m_j)}\rangle_C$.\\

The mutual information $I_{ij}$ is presented next to the single orbital entropy in \cref{fig:S1_I}.
In general, the correlation effects increase towards the dissociation limit.
This effect is connected to the choice of canonical, i.e.~delocalized, orbitals. A reversed effect has been observed for localized orbitals in \ch{Be6} rings \cite{Fertitta-2014}, where entropies are large at equilibrium distance and small at the separated atom limit.
We observe that most correlation is between the $3\sigma$ and $1\pi$ bonding/anti-bonding (gerade/ungerade parity) pairs: $1\pi_u^\prime + 1\pi_g^\prime$ and $1\pi_u^{\prime\prime} + 1\pi_g^{\prime\prime}$ (blue) and $3\sigma_g + 3\sigma_u$ (green). As before, the curve for the $3\sigma$ pair is shifted towards larger bond distances, but ends up at the same dissociation limit. Smaller contribution arise from correlations between the degenerate $1\pi$ pairs of same parity, with a maximum at $R\approx 1.90\ \rm \AA$: $1\pi_u^\prime + 1\pi_u^{\prime\prime}$ and $1\pi_g^\prime + 1\pi_g^{\prime\prime}$ (red).
Interestingly, these two pairs have almost the same mutual informations, even though the pairs have different energies and occupations. This follows from the very symmetric behavior of the $3\sigma$ and $1\pi$ orbitals, as discussed above for the one orbital entropies.\\

Slightly smaller and with the maximum slightly shifted to a larger distance ($R\approx2.00\ \rm \AA$) is the mutual information for the $1\pi_u^\prime + 1\pi_g^{\prime\prime}$ and $1\pi_g^\prime + 1\pi_u^{\prime\prime}$ pairs (cyan). Both lines end up in the same dissociation limit as the different combination of $3\sigma$ and $1\pi$ orbitals, since they evolve towards the degenerate $2p$ orbitals of the isolated fragments.\\

\begin{table}
	\centering
	\caption{Assignment of labels to characteristic eigenvectors $\phi_{ij,\alpha}$ of the two-orbital density matrix. Further possible eigenvectors are the basis vectors itself, which are indicated by their corresponding label directly.}
	\begin{tabular}{ccc}
		$\phi_{ij,\alpha}$ & $\langle S^2\rangle$ & label \\
		\hline
		$\frac{1}{\sqrt{2}} (|-,\uparrow\downarrow\rangle + |\uparrow\downarrow,-\rangle)$ & 0 & $|\uparrow\downarrow,-\rangle_+$ \\
		$\frac{1}{\sqrt{2}} (|-,\uparrow\downarrow\rangle - |\uparrow\downarrow-\rangle)$ & 0 & $|\uparrow\downarrow,-\rangle_-$ \\
		$\frac{1}{\sqrt{2}} (|\uparrow,\downarrow\rangle - |\downarrow,\uparrow\rangle)$ & 0 & singlet \\
		$\frac{1}{\sqrt{2}} (|\uparrow,\downarrow\rangle + |\downarrow,\uparrow\rangle)$ & 2 & \multirow{3}{*}{$\Bigg\}$ triplet} \\
		$|\uparrow,\uparrow\rangle$ & 2 & \\
		$|\downarrow,\downarrow\rangle$ & 2 & \\
	\end{tabular}
	\label{tab:phi2_labels}
\end{table}

For the diagonalized two-orbital reduced density matrix $\rho_{ij}$ and the connected part of the generalized correlation functions a lot of data is obtained. To simplify the discussion here, we restrict ourselves to the two orbital pairs with highest mutual information: $1\pi_u^\prime+1\pi_g^\prime$ and $3\sigma_g+3\sigma_u$. For further orbital pairs we refer to the Supplementary Information.\\

For a compact graphical representation of the diagonalized two-orbital reduced density matrix, only eigenvalues with largest contribution are considered. Characteristic eigenvectors which are constant over the internuclear distance $R$ are labeled according to \cref{tab:phi2_labels}, others are represented by a plot of their non-zero coefficients $c_{ij,\alpha}$. Please note, that a \enquote{triplet} contribution in a given orbital pair does not necessarily indicate a triplet character of the wave function, as the spin states of the remaining orbitals in a given configuration contribute to the total spin as well. Spin states higher than triplet cannot be formed from just two orbitals.\\

\begin{figure}
	\centering
	\includegraphics[width=\linewidth]{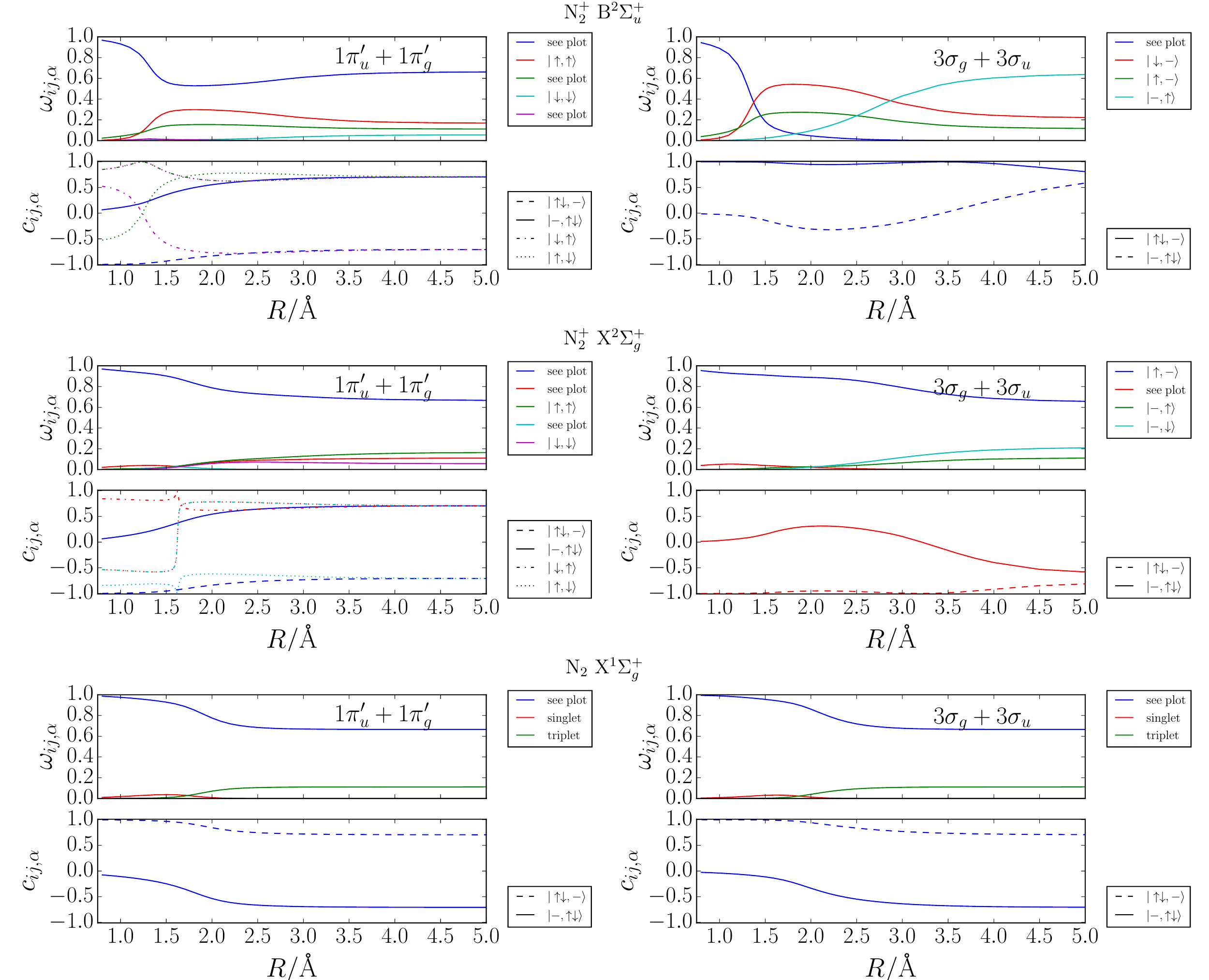}
	\caption{Largest eigenvalues $\omega_{ij,\alpha}$ and their corresponding eigenvectors $\phi_{ij,\alpha}=\sum_\alpha c_{ij,\alpha}|\phi_{\alpha_i}\rangle|\phi_{\alpha_j}\rangle$ of the two-orbital reduced density matrices, plotted over the internuclear distance $R$.
	Selected orbital pairs are $1\pi_u^\prime + 1\pi_g^\prime$ and $3\sigma_g + 3\sigma_u$, left and right columns respectively.
	The electronic states from top to bottom are \ch{N2+} $\mathrm{B}\prescript{1}{}\Sigma_u^+$, \ch{N2+} $\mathrm{X}\prescript{1}{}\Sigma_g^+$ and \ch{N2} $\mathrm{X}\prescript{1}{}\Sigma_g^+$.
	Labels for eigenvectors which do not change with respect to $R$ are assigned according to \cref{tab:phi2_labels}, $s$ indicates $\uparrow$ and $\downarrow$ yield same results. For eigenvectors depending on $R$ their coefficients are plotted.}
	\label{fig:phi2}
\end{figure}

The diagonalized two-orbital reduced density matrix for the \ch{N2} ground state is presented in \cref{fig:phi2} (bottom row).
Both orbital pairs show similar occupations, starting with high double occupancy of the energetically lower, bonding orbital for small $R$ and evolving towards the superposition $\frac{1}{\sqrt{2}} (|-,\uparrow\downarrow\rangle - |\uparrow\downarrow,-\rangle) = |\uparrow\downarrow,-\rangle_-$ at dissociation limit.
During bond breaking they gain some single occupations character, as indicated by the singlet and triplet eigenstates (red and green respectively), as we could already observe for eigenvalues of the one-orbital density matrix (cf \cref{fig:w1}). However, this time we gain some information on their relative spin (up/down). First, during bond breaking, we get some singlet character, which then quickly changes to a triplet contribution towards the dissociate limit, accounting for the quadruplet spin state of the $\mathrm{N} (\mathrm{\prescript{4}{}S}_u)$ fragments.
However, contributions to electron correlation are dominated by configurations where one orbital is doubly occupied while the other one is empty. \\

\begin{figure}
	\centering
	\includegraphics[height=0.89\textheight,width=\linewidth,keepaspectratio]{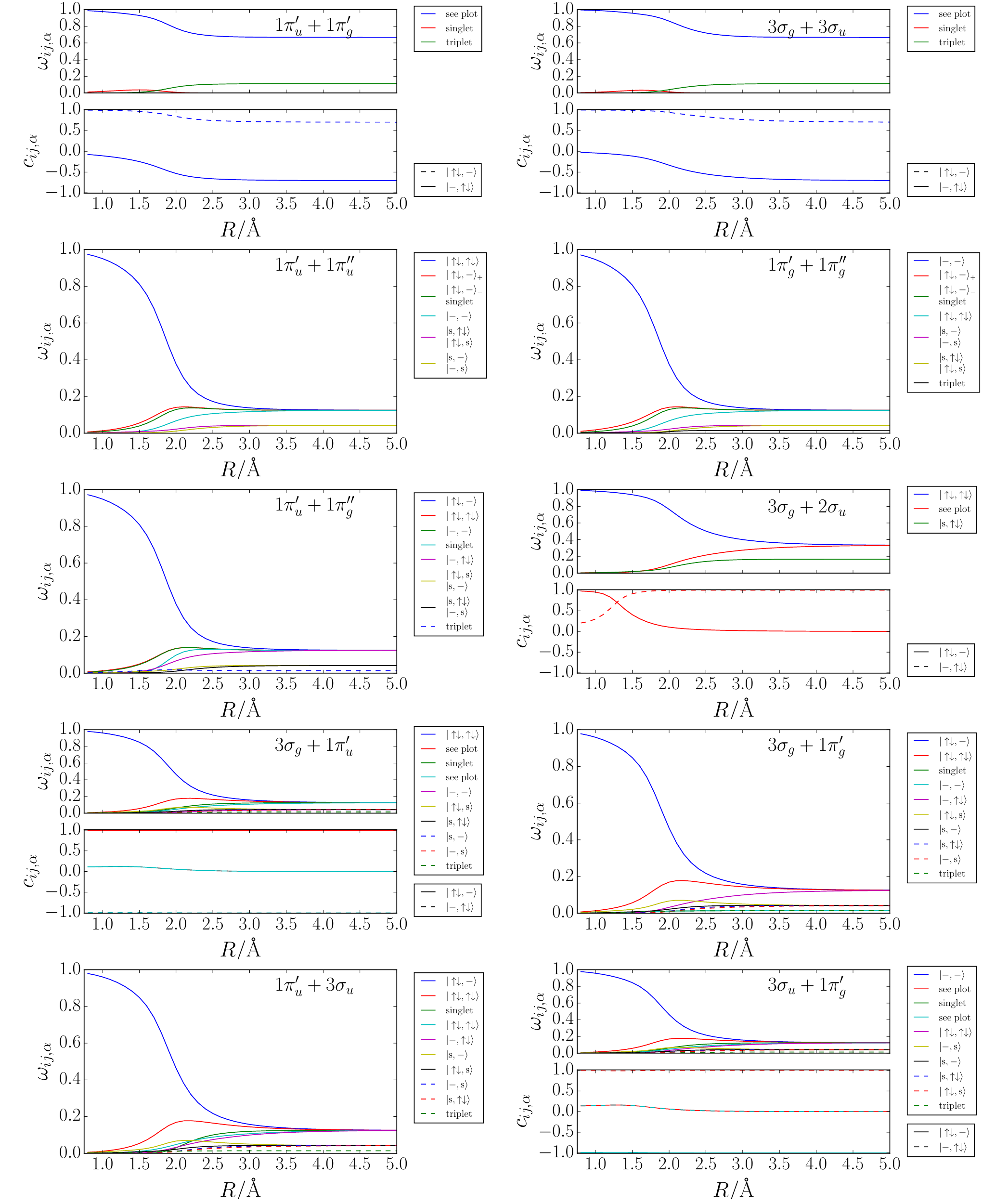}
	\caption{Largest eigenvalues $\omega_{ij,\alpha}$ and their corresponding eigenvectors $\phi_{ij,\alpha}=\sum_\alpha c_{ij,\alpha}|\phi_{\alpha_i}\rangle|\phi_{\alpha_j}\rangle$ of the two-orbital reduced density matrices
	as a function of internuclear distance $R$ for the \ch{N2} $\mathrm{X}\prescript{2}{}\Sigma_g^+$ state. Labels for eigenvectors which do not change with respect to $R$ are assigned according to \cref{tab:phi2_labels}, $s$ indicates a single electron which can have up ($\uparrow$) or down ($\downarrow$) spin. For eigenvectors depending on $R$ their coefficients are plotted.}
	\label{fig:phi2_XN2}
\end{figure}

The remaining pairs (\cref{fig:phi2_XN2}) have one major eigenstate for small $R$ as well, allowing to easily identity the leading HF configuration at equilibrium distance. As electron correlation increases towards the dissociate limit, we can identify a number of many degenerate eigenstates representing all kinds of different occupations.\\

Static, non-dynamic and dynamic correlations can be characterized by the value of the mutual information $I_{ij}$ in certain extend ($0<I_{ij}<2\times\ln(16)\approx2.77$). Large $I_{ij}$ corresponds to static, intermediate values to non-dynamic and small values to dynamic correlations \cite{Boguslawski-2012, Ehlers-2015}.
Taking a look at \cref{fig:S1_I} for orbitals pair $1\pi_u^\prime+1\pi_g^\prime$, the mutual information changes from almost zero to $\approx 1.6$, correspondingly in \cref{fig:phi2_XN2} for large $R$ we have 2 finite eigenvalues and around equilibrium distance only one eigenvalue remains finite and, being close to 1, indicating a pure state for the two-orbital subsystem.
In contrast to this, for $1\pi_u^\prime+1\pi_u^{\prime\prime}$ the mutual information remains small for all $R$ values, corresponding to dynamic and non-dynamic correlations. In \cref{fig:phi2_XN2} for large $R$ we have 2 highly degenerate levels (showing that the 2 orbital subsystem is in a highly mixed state) with finite but small values and for smaller distances the degeneracy is lifted until around equilibrium distance only one eigenvalue is approximately 1, again indicating a pure state for the two-orbital subsystem.
Therefore we can identify static correlations for the pairs $1\pi_u^\prime+1\pi_g^\prime$, $1\pi_u^{\prime\prime}+1\pi_g^{\prime\prime}$ and $3\sigma_g+3\sigma_u$ shown in \cref{fig:phi2}, and dynamic correlations for all remaining pairs.\\

\begin{figure}
	\centering
	\includegraphics[width=\linewidth]{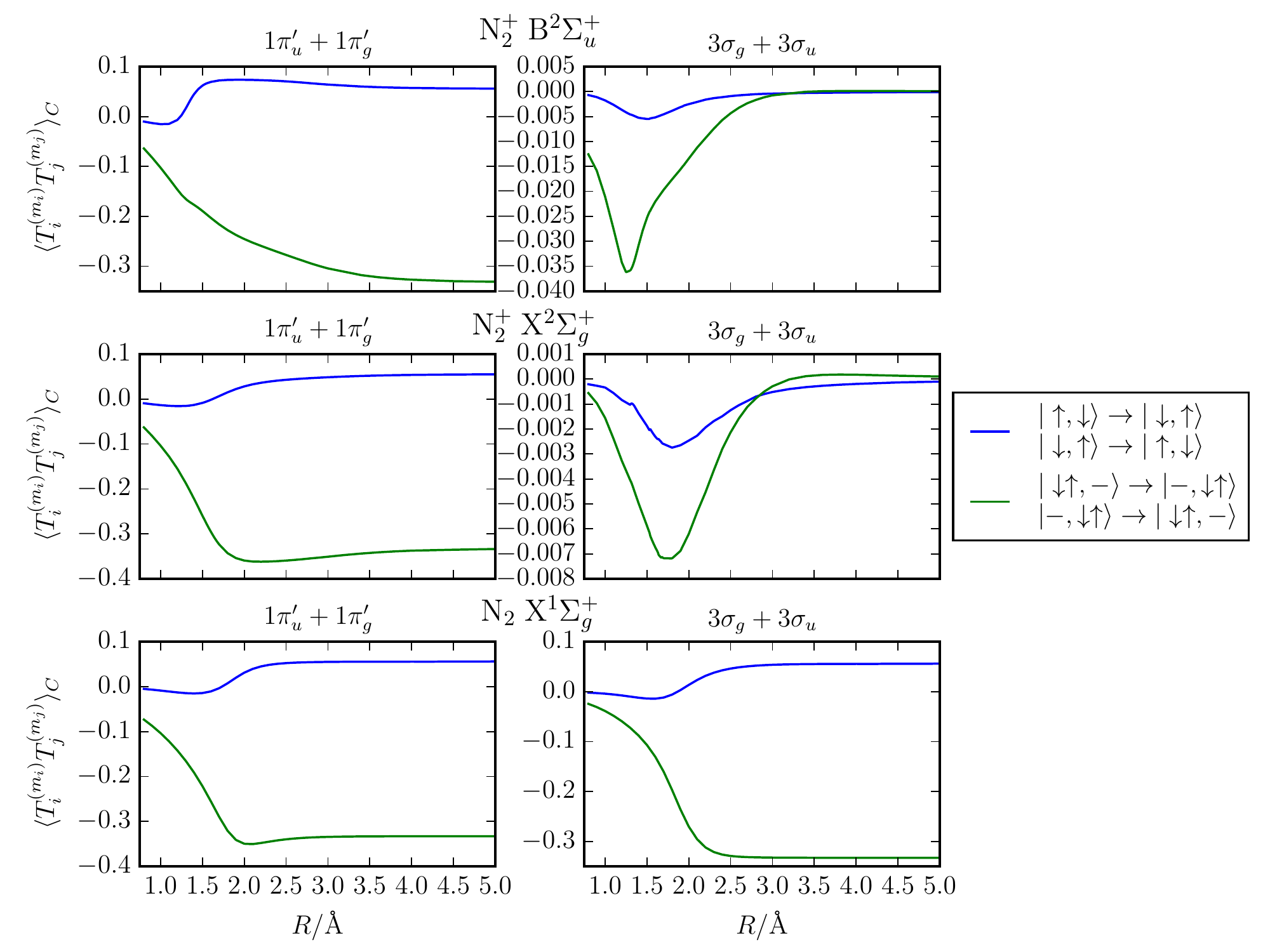}
	\caption{Correlations functions $\langle T_i^{(m_i)} T_j^{(m_j)}\rangle_C(R)$ as a function of internuclear distance $R$ for the two orbital pairs $1\pi_u^\prime + 1\pi_g^\prime$ and $3\sigma_g + 3\sigma_u$, left and right columns respectively. Only the connected contributions (cf. \cref{eq:correl_C}) where $\max(|\langle T_i^{(m_i)} T_j^{(m_j)}\rangle_C(R)|)>10^{-6}$ are shown.
	The electronic states from top to bottom are \ch{N2+} $\mathrm{B}\prescript{1}{}\Sigma_u^+$, \ch{N2+} $\mathrm{X}\prescript{1}{}\Sigma_g^+$ and \ch{N2} $\mathrm{X}\prescript{1}{}\Sigma_g^+$.}
	\label{fig:correl}
\end{figure}

In \cref{fig:correl} (bottom row) the connected contributions of the generalized correlation functions $\langle T_i^{(m_i)} T_j^{(m_j)}\rangle_C$ are presented, for the same two orbital pairs as for $\phi_{ij}$.
Note, that disconnected contributions are subtracted out. For example the major contributions of the diagonal \mbox{$|\downarrow\uparrow,-\rangle\rightarrow|\downarrow\uparrow,-\rangle$} transition operator for the $1\pi_u^\prime+1\pi_g^\prime$ pair, mostly arising from the HF configuration, would be clearly visible in $\langle T_i^{(m_i)} T_j^{(m_j)}\rangle$ but does not show up in $\langle T_i^{(m_i)} T_j^{(m_j)}\rangle_C$.
The diagonal elements, e.g. $|-,\downarrow\uparrow\rangle\rightarrow|-,\downarrow\uparrow\rangle$ and $|\downarrow\uparrow,-\rangle\rightarrow|\downarrow\uparrow,-\rangle$ provide information on the occupation of the orbital pair. While the off-diagonal elements shows the resonance between the different basis configurations. \\

As above for $\phi_{ij}$, we can observe increasing correlation effects towards the dissociation limit. The $1\pi_u^\prime+1\pi_g^\prime$ and $3\sigma_g+3\sigma_u$ (top row) show largest magnitudes (about 3 times larger than for other pairs) and, as expected, are very similar to each other. In accordance with the above identification of static and dynamic correlation effects, the largest correlation functions are the $|\downarrow\uparrow,-\rangle\rightarrow|-,\downarrow\uparrow\rangle$ (and vice versa) resonances.
In general only few and small contributions are observable for around equilibrium distance, but are rapidly increasing during bond breaking as correlation effects increase.\\

The only other appearing off-diagonal transition operator are $|\uparrow,\downarrow\rangle\rightarrow|\downarrow,\uparrow\rangle$ and $|\downarrow,\uparrow\rangle\rightarrow|\uparrow,\downarrow\rangle$. These have a positive sign, again matching the triplet character observed in $\phi_{ij}$ above.\\

Additionally, some emerging patterns in the correlation functions plots are very similar, e.g. $1\pi_u^\prime + 1\pi_g^\prime$ and $3\sigma_g+3\sigma_u$ show only minor differences. Both pairs come from the $2p$ shell of the atomic fragments and combine orbitals of different parity ($g/u$) but otherwise same symmetry. Similar patterns can also be observed for $1\pi_u^\prime+1\pi_u^{\prime\prime}$ and $1\pi_u^\prime+1\pi_g^{\prime\prime}$ as well as $1\pi_u^\prime+3\sigma_u$ and $3\sigma_u+1\pi_g^\prime$ (cf. Fig. S1).\\


\subsubsection[N2+ ground state]{\ch{N2+} $\mathrm{X}\prescript{2}{}\Sigma_g^+$ ground state}

Next, we investigate the \ch{N2+} $\mathrm{X}\prescript{2}{}\Sigma_g^+$ doublet ground state (middle row in \cref{fig:w1,fig:S1_I,fig:phi2,fig:correl}), which shares some similarities with the \ch{N2} singlet ground state: the entropy increases during dissociation and the entanglement of the $1\pi$ orbitals remains about the same. \\

Major differences are observed for the $3\sigma_{g/u}$ orbitals, which are no longer doubly occupied. This represents the electron hole arising for the positive charge of the cation. Both orbitals remain mainly in a singly occupied state, even at the dissociation limit.
Accordingly the $3\sigma_g$ and $3\sigma_u$ entropies drop down by about a factor of 2 compared to the \ch{N2} $\mathrm{X}\prescript{1}{}\Sigma_g^+$ ground state.
Instead $2\sigma_{g/u}$ entropies are increased, as the $3\sigma_g$ orbital, being close in energy, opened up, allowing for possible excitations. This effect is larger for the $2\sigma_u$, which is much closer in energy to the $3\sigma_g$ (cf. \cref{fig:orbenergy}). Towards the dissociation limit the $2\sigma_{g/u}$ orbitals show the same entropy, as they become degenerate.\\

For the calculation the total spin of the doublet state was chosen to be in the spin up state.
A closer look to the single spin occupations (cf \cref{fig:w1}) reveals that the electron hole leaves a pronounced up-spin character in the $3\sigma_g$ orbital, corresponding to the total spin of the electronic state. Similarly the $1\pi_{g/u}$ have more up-spin than down-spin character, while the $3\sigma_u$ orbital surprises with down-spin character.\\

In the mutual information (middle right panel in \cref{fig:S1_I}) the $1\pi$ bonding/anti-bonding pairs (blue) are very similar to the \ch{N2} ground state. However, the $3\sigma_g + 3\sigma_u$ reduces to about one third, due to the electron hole being located in the $3\sigma_g$ orbital (cf. \cref{fig:w1}). In turn the $1\pi_u^\prime + 1\pi_u^{\prime\prime}$ (red) and $1\pi_g^\prime + 1\pi_g^{\prime\prime}$ (black) mutual information increase by a factor of two for small and intermediate $R$, but remain about the same at the dissociation limit. Additionally, those two orbital pairs are no longer close in their mutual information, but show a very small deviation, since the occupations of the $3\sigma$ orbitals are not symmetric anymore.
Furthermore, we observe a number of smaller additional contributions (thin black lines). Thus we observe higher electron correlations.\\

Accordingly, the diagonalized two-orbital reduced density matrices in \cref{fig:phi2} (middle row) share some similarities with the \ch{N2} ground state as well (see Fig. S2), e.g. $1\pi_g^\prime+1\pi_g^{\prime\prime}$ and $1\pi_u^\prime+1\pi_u^{\prime\prime}$. But they also show some interesting new features:
For example the $1\pi_u^\prime+1\pi_g^\prime$ pair follows a similar trend with respect to the singlet and triplet contributions. However, the previously degenerate triplet components now split towards the dissociation limit: while the $\frac{1}{\sqrt{2}}(|\downarrow,\uparrow\rangle+|\uparrow,\downarrow\rangle)$ component remains about the same, the $|\uparrow,\uparrow\rangle$ component increases at the cost of the $|\downarrow,\downarrow\rangle$ component, due to the overall spin doublet character of the $\mathrm{X}\prescript{2}{}\Sigma_g^+$ state. Accordingly, for $R < 1.6\ \rm \AA$, the small singlet contribution $\frac{1}{\sqrt{2}}(|\uparrow,\downarrow\rangle-|\downarrow,\uparrow\rangle)$, as observed for \ch{N2}, now slightly changed to around $0.55|\uparrow,\downarrow\rangle-0.84|\downarrow,\uparrow\rangle$ which contributes $\langle S^2 \rangle=0.08$ to the total spin.\\

Other major changes occur for pairs connected to the electron hole, most importantly for $3\sigma_g+3\sigma_u$, where most singlet and triplet contributions are replaced by contributions that account for the doublet state.
Furthermore, as already observed for $\omega_i$ above, the $3\sigma_g$ orbital is dominated by up-spin character while $3\sigma_u$ has down-spin character.\\

The connected part of the generalized correlations in \cref{fig:correl} (middle row) are almost identical to the \ch{N2} case for the $1\pi$ orbital pairs. Pairs including $3\sigma$ orbitals are smaller in magnitude for small $R$ but similar at dissociation limit, though the transition operators are different due to different occupations in the corresponding configurations. The similar patterns, as observed for the \ch{N2} ground state, are still observable but less pronounced.\\

\subsubsection[N2+ excited state]{\ch{N2+} $\mathrm{B}\prescript{2}{}\Sigma_u^+$ excited state}

A much more complex picture is observed for the \ch{N2+} $\mathrm{B}\prescript{2}{}\Sigma_u^+$ excited state (upper row in \cref{fig:w1,fig:S1_I,fig:phi2,fig:correl}).
Biggest differences are found during bond breaking, i.e. from $R\approx 1.1$ to $R\approx 4\ \rm\AA$, which relates to the change of the leading configuration as already apparent in the CI vector (cf. \cref{fig:coeffs}). The situation at the dissociation limit is almost the same as for the \ch{N2+} ground state, since both states dissociate into the same atomic fragments (cf. \cref{fig:pes_all_states}).\\

Starting the discussion again with the spin state probability (\cref{fig:w1}), we see some steep changes just after the equilibrium distance ($R_e=1.1\ \rm\AA$): The $2\sigma_u$ orbital goes from a high up-spin occupation (the corresponding curve overlaps to a large extend with the $3\sigma_g$ double occupation) to double occupation, while the $3\sigma_g$ and $1\pi_u$ orbitals evolve from a doubly to a mixture of doubly and singly occupied states. The same effect is observed in \cref{fig:coeffs}, where the leading configuration switches somewhere shortly before $R=1.5\ \rm\AA$.
Accordingly, we see a peak in the $2\sigma_u$ orbital entropy and fluctuating values for the $3\sigma$ and $1\pi$ orbital entropies (\cref{fig:S1_I}).\\

A closer look to the single occupations reveals high down-spin character in the bonding $3\sigma_g$ orbital, accompanied by up-spin character in the $3\sigma_u$ and all $1\pi$ orbitals. Overall a total up-spin doublet state is retained, just as for the \ch{N2+} ground state. At dissociation limit both states show the same spin occupations, just with the roles of the, here becoming degenerate, $3\sigma_{g/u}$ orbitals exchanged.\\

The mutual information (top row in \cref{fig:S1_I}) is dominated by the $1\pi_u^\prime + 1\pi_g^\prime$ pair, similar to the \ch{N2+} $\mathrm{X}\prescript{2}{}\Sigma_u^+$ ground state. Secondary contributions are quite different for intermediate $R$ (during bond breaking): The $3\sigma_g$ has increased entropy and is entangled with many different orbitals.
Please note, that the red line represents three orbital pairs, which have only small differences and are represented as one for better visibility.\\

The diagonalized two-orbital density matrices (top row in \cref{fig:phi2}) reflect the more complex occupations patterns around bond breaking as well. The triplet splitting in the $1\pi_u^\prime+1\pi_g^\prime$ is much larger as for the \ch{N2+} ground state and has maximum around $1.8\ \rm\AA$.
In the $3\sigma_g+3\sigma_u$ pair we can see how the $3\sigma_g$ starts for short distances being mainly doubly occupied, then becomes singly occupied, with down-spin contribution being about twice as large as up-spin, and ends up with a mixture of single and empty occupation states at the dissociation limit. At the same time the $3\sigma_u$ evolves from an empty spin occupation towards single occupation.
The correlation functions are quite similar to the ones of the \ch{N2+} ground state.\\

\section{Summary and Discussion}
\label{sec:5_summary}

Highly accurate potential energy surfaces for \ch{N2} and \ch{N2+} including excited states have been calculated, both on the MRCI-SD and DMRG level demonstrating the current capabilities. Furthermore we showed how the application of quantum information theory can give insights about the electronic structure of strongly correlated system. Obviously, to describe electron correlation in the \ch{N2} ground state $\mathrm{X}\prescript{1}{}\Sigma_g^+$, the $3\sigma$ and $1\pi$ orbitals are most important. While the $2\sigma$ orbitals only play a minor role, but are increasingly important for the energetically higher lying \ch{N2+} ground state $\mathrm{X}\prescript{2}{}\Sigma_g^+$ and excited state $\mathrm{B}\prescript{2}{}\Sigma_u^+$.\\

By comparing the QIT results between different states and charged species it is easily possible to locate the electron hole as differentiating different spin multiplicities.
We could show that orbital correlations are primarily between pairs overlapping in space and differing in symmetry only by gerade/ungerade parity.\\

Furthermore we can use the diagonalized two-orbital reduced density matrices $\rho_{ij}$ in connection with the mutual informations to classify dynamic, non-dynamic and static correlations and the corresponding relevant configurations. Applying such an analysis to strongly correlated and large systems, may lead to truncation schemes neglecting all dynamic correlations, by not only selecting an active space, but further restricting systematically the occupations of these orbitals.\\

In general the QIT quantities have a rather simple and ordered structure for the \ch{N2} ground state, but more complex patterns emerge when going to higher excited systems and states which are more strongly correlated.

\begin{acknowledgments}
	We gratefully acknowledge financial support from the International Max Planck Research School ``Functional Interfaces in Physics and Chemistry''. The high performance computing facilities of the Freie Universität Berlin (ZEDAT) are acknowledged for computing time.
	O.L. acknowledges support in part by the Hungarian Research Fund (OTKA) through Grant No NN110360, the National Research, Development and Innovation Office (NKFIH) through Grant No K120569) and the Hungarian Quantum Technology (HunQtech) through Grant No 2017-1.2.1-NKP-2017-00001.
\end{acknowledgments}

\bibliography{DMRG_QIT.bib}

\end{document}